\documentclass[10pt]{article}

\usepackage{amsmath}
\usepackage{amssymb}

\usepackage{graphicx}

\usepackage{cite}

\usepackage{color}
\newcommand{\rev}[1]{\textcolor{black}{#1}}
\newcommand{\revtwo}[1]{\textcolor{black}{#1}}

\usepackage[labelfont=bf,labelsep=period,justification=raggedright]{caption}

\makeatletter
\renewcommand{\@biblabel}[1]{\quad#1.}
\makeatother

\date{}

\begin{document}

\begin{flushleft}
{\Large
\textbf{\revtwo{Task-Based} Core-Periphery Organization of Human Brain Dynamics}
}
\\

Danielle S. Bassett$^{1,2,\ast}$,
Nicholas F. Wymbs$^{5}$,
M. Puck Rombach$^{3,4}$,
Mason A. Porter$^{3,4}$,
Peter J. Mucha$^{6,7}$,
Scott T. Grafton$^{5}$
\\
\bf{1} Department of Physics, University of California, Santa Barbara, CA 93106, USA
\\
\bf{2} Sage Center for the Study of the Mind, University of California, Santa Barbara, CA 93106, USA
\\
\bf{3} Oxford Centre for Industrial and Applied Mathematics, Mathematical Institute, University of Oxford, Oxford OX1 3LB, UK
\\
\bf{4} CABDyN Complexity Centre, University of Oxford, Oxford, OX1 1HP, UK
\\
\bf{5} Department of Psychological and Brain Sciences and UCSB Brain Imaging Center, University of California, Santa Barbara, CA 93106, USA
\\
\bf{6} Carolina Center for Interdisciplinary Applied Mathematics, Department of Mathematics, University of North Carolina, Chapel Hill, NC 27599, USA
\\
\bf{7} Institute for Advanced Materials, Nanoscience \& Technology, University of North Carolina, Chapel Hill, NC 27599, USA
\\
$\ast$ E-mail: Corresponding dbassett@physics.ucsb.edu
\end{flushleft}




\section*{Abstract}

As a person learns a new skill, distinct synapses, brain regions, and circuits are engaged and change over time. In this paper, we \rev{develop} methods to examine patterns of correlated activity across a large set of brain regions. Our goal is to identify properties that enable robust learning of a motor skill. We measure brain activity during motor sequencing and characterize network properties based on coherent activity between brain regions. Using recently developed algorithms to detect time-evolving communities, we find that the complex reconfiguration patterns of the brain's putative functional modules that control learning can be described parsimoniously by the combined presence of a relatively stiff \revtwo{temporal} \emph{core} that is composed primarily of sensorimotor and visual regions whose connectivity changes little in time and a flexible \revtwo{temporal} \emph{periphery} that is composed primarily of multimodal association regions whose connectivity changes frequently. The separation between \revtwo{temporal} core and periphery changes over the course of training and, importantly, is a good predictor of individual differences in learning success. The core of dynamically stiff regions exhibits dense connectivity, which is consistent with notions of core-periphery organization established previously in social networks. Our results demonstrate that core-periphery organization provides \rev{an insightful way} to understand how putative functional modules are linked. This, in turn, enables the prediction of fundamental human capacities, including the production of complex goal-directed behavior.





\section*{Author Summary}

When someone learns a new skill, his/her brain dynamically alters individual synapses, regional activity, and larger-scale circuits. In this paper, we capture some of these dynamics by measuring and characterizing patterns of coherent brain activity during the learning of a motor skill. We extract time-evolving communities from these patterns and find that a \revtwo{temporal} core that is composed primarily of primary sensorimotor and visual regions reconfigures little over time, whereas a periphery that is composed primarily of multimodal association regions reconfigures frequently. The core consists of densely connected nodes, and the periphery consists of sparsely connected nodes. Individual participants with a larger separation between core and periphery learn better in subsequent training sessions than individuals with a smaller separation. Conceptually, core-periphery organization provides a framework in which to understand how putative functional modules are linked. This, in turn, enables the prediction of fundamental human capacities, including the production of complex goal-directed behavior.

%


\section*{Introduction}

Cohesive structures have long been thought to play an important role in information processing in the human brain \cite{Saalmann2012}. At the small scale of individual neurons, temporally coherent activity supports information transfer between cells \cite{Josic2009}. At a much larger scale, simultaneously active cortical areas form functional systems that enable behavior \cite{Saalmann2012}. However, the question of precisely what type of cohesive organization is present between the constituents of brain systems---\rev{especially at larger scales}---has been steeped in controversy \cite{Op2008,Plaut1995}. Although low-frequency interactions between pairs of brain areas are easy to measure, the simultaneous characterization of dynamic interactions across the entire human brain remained challenging until recent applications of network theory to neuroimaging data \cite{Bullmore2011}. These efforts have led to enormous insights, including the establishment of relationships between stationary functional brain network configuration and intelligence \cite{Langer2012} as well as relationships between altered brain network organization and disease \cite{Bassett2009b}. In this paper, we extend this approach to a non-stationary situation: the change of network activity across the brain as a new skill is acquired.

Acquisition of new motor skills \rev{alters brain activity across spatial scales.  At the level of individual neurons, this induces changes in firing behavior in the motor cortex \cite{Matsuzaka2007}. At the level of large-scale areas, this induces changes in the interactions between primary motor cortex and premotor areas, and these changes can influence the amount of learning \cite{Wymbs2009}.} Previous studies have demonstrated that pairwise interactions between some of these premotor regions, as measured by the magnitude of coherence \rev{between low-frequency blood-oxygen-level-dependent (BOLD) signals}, strengthen with practice \cite{Sun2007}. Furthermore, complex contributions by non-motor systems such as prefrontal cortex are involved in the strategic control of behavior during learning \cite{Mushiake2006}. These findings reveal some of the changes in local circuits that occur with learning.  However, there remains no global assessment of changes in brain networks as a result of learning. In this paper, we seek to find cohesive structures in global brain networks that capture dynamics that are particularly relevant for characterizing skill learning that takes place over the relatively long time scales of minutes to hours of practice.

\begin{figure}[h]
\begin{center}
\includegraphics[width=.65\linewidth]{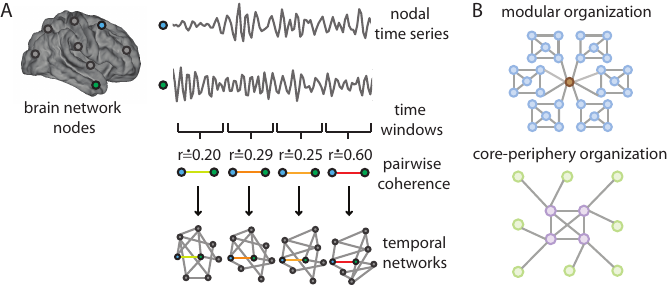}
    \caption{\label{fig:meth1} \textbf{Network Organization of Human Brain Dynamics}. \emph{(A)} Temporal Networks of the Human Brain. We parcellate the brain into anatomical regions that can be represented as nodes in a network, and we use the coherence between functional Magnetic Resonance Imaging (fMRI) time series of each pair of nodes over a time window to determine the weight of the network edge connecting those nodes. We determine these weights separately using approximately 10 non-overlapping time windows of 2--3 min duration and thereby construct temporal networks that represent the dynamical functional connectivity in the brain. \emph{(B)} Cohesive Mesoscale Structures. \emph{(top)} An example of a network with a modular organization in which high-degree nodes (brown) are often found in the center of modules or bridging distinct modules that are composed mostly of low-degree nodes (blue). \emph{(bottom)} A network with a core-periphery organization in which nodes in the core (purple) are more densely connected with one another than nodes in the periphery are with one another (green).
    }
    \end{center}
\end{figure}

To address these issues, we extract a set of functional networks from task-based functional Magnetic Resonance Imaging (fMRI) time series that describe functional connectivity between brain regions. We probe the dynamics of these putative interactions by subdividing time series into discrete time intervals (of approximately two minutes in duration; see Fig.~\ref{fig:meth1}A) during the acquisition of a simple motor skill. Subjects learned a set of 10-element motor sequences similar to piano arpeggios by practicing for at least 30 days during a 6-week period. The depth of training was manipulated so that 2 sequences were extensively practiced (EXT), 2 sequences were moderately practiced (MOD), and 2 sequences were minimally practiced (MIN) on each day. In addition, subjects performed blocks of all of the sequences during fMRI scanning on approximately days 1, 14, 28, and 42 of practice. Using the fMRI time series, we extract functional networks representing the coherence between 112 cortical and subcortical areas for each sequence block.

To characterize brain dynamics, \rev{we represent sets of functional networks as multilayer brain networks and we identify putative functional modules---i.e., groups of brain regions that exhibit similar BOLD time courses---in each 2--3 minute time window. Such cohesive groups of nodes are called ``communities'' in the network-science literature \cite{Porter2009,Fortunato2010}, and they suggest that different sets of brain regions might be related to one another functionally either through direct anatomical connections or through indirect activation by an external stimulus. A community of brain regions might code for a different function (e.g., visual processing, motor performance, or cognitive control), or it might engage in the same function using a distinct processing stream. Characterizing changes in community structure thus makes it possible to map meaningful dynamic patterns of functional connectivity that relate to changes in cognitive function (e.g., learning).}

\rev{We employ computational tools for dynamic community detection \cite{Mucha2010,Bassett2012b} for multilayer representations of temporal networks \cite{Holme2011} and summarize our findings using diagnostics that quantify three properties of community structure. (See Materials and Methods for their definitions and Ref.~\cite{Bassett2011b} for evidence supporting the utility of these diagnostics in capturing changes in brain dynamics over 3 days of learning.) To measure the strength of functional modularization in the brain and quantify the extent of compartmentalization of putative functional modules, we maximize a quality function called \emph{multilayer modularity} $Q$ to obtain a partition of the brain
into communities. (The associated maximum value of $Q$ is known as the \emph{maximum modularity}.) A high value of $Q$ indicates that the pattern of functional connectivity in the brain can be clustered sensibly into distinct communities of brain regions that exhibit similar time courses. We also compute the number $n$ of communities (i.e., putative functional modules) in partitions of the multilayer networks. A large value of $n$ indicates that there are a large number of distinct temporal profiles in BOLD activations in the brain. To measure the temporal variability of community structure, we compute the \emph{flexibility} $f_i$ of each region $i$, as this quantifies the frequency that a brain region $i$ changes its allegiances to network communities over time. A high value of flexibility indicates that a region often changes community affiliation.}


Our results demonstrate that the temporal evolution of community structure is modulated strongly by the depth of training (as reflected in the total number of practiced trials). We also show that the temporal variability of module allegiance varies across brain regions. Sensorimotor and visual cortices form the bulk of a relatively stiff \revtwo{temporal} core in which module affiliations change little over a scanning session, whereas multimodal association areas form the bulk of a relatively flexible \revtwo{temporal} periphery in which module affiliations change frequently. The separation between the temporal core and temporal periphery predicts individual differences in extended learning. We combine these methods for identifying a temporal core and periphery with a notion of core-periphery organization that originated in the social sciences \cite{Borgatti1999} to show that the organizational structure of \revtwo{functional} networks in 2--3 minute time windows correlates with the organizational structure of the brain's temporal evolution: densely connected regions in individual time windows tend to exhibit little change in module allegiance over time, whereas weakly connected regions tend to exhibit significant changes. Taken together, our results suggest that core-periphery organization is a critical property that is as important as modularity for understanding and predicting cognition and behavior (see Fig.~\ref{fig:meth1}B).


\section*{Results}


\subsection*{Dynamic Community Structure Changes with Learning}

\begin{figure}
\begin{center}
\includegraphics[width=1\linewidth]{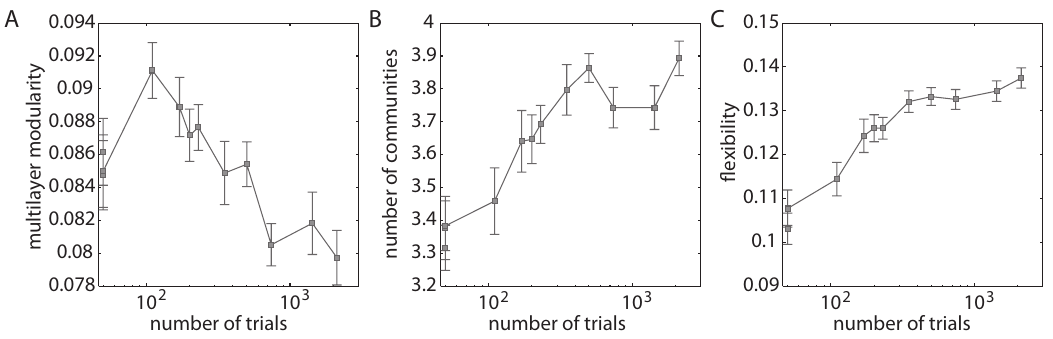}
    \caption[]{\label{fig:mnf} \textbf{Dynamic Network Diagnostics Change with Learning}. \emph{(A)} Multilayer modularity, \emph{(B)} number of communities, and \emph{(C)} mean flexibility calculated as a function of the number of trials completed after a scanning session (see Table \ref{experiment0} for the relationship between the number of trials practiced and training duration and intensity). We average the values for each diagnostic over the 100 multilayer modularity optimizations, and we average flexibility over the 112 brain regions (in addition to averaging over the 100 optimizations per subject). Error bars indicate the standard error of the mean over participants.
       }
    \end{center}
\end{figure}

Community structure changes with the number of trials practiced, independent of when the practice occurred in the 6 weeks. In Fig.~\ref{fig:mnf}, we show multilayer modularity ($Q$, a measure of the quality of a partition into communities), the number of communities, and mean flexibility ($F=\frac{1}{N}\sum_{i=1}^N f_{i}$, a measure of the temporal variability in module allegiance) as a function of the number of training trials completed after a scanning session. \rev{See Materials and Methods for the definitions.} After an initial increase from 50 to 200 trials practiced, multilayer modularity decreases with an increase in the number of trials practiced, suggesting that community structure in functional brain networks becomes less pronounced with learning. Both the number of communities and the flexibility of community structure increase with the number of trials practiced, which is consistent with an increased specificity of functional connectivity patterns with extended learning.


\subsection*{Temporal Core-Periphery Organization}

\begin{figure}
\begin{center}
\includegraphics[width=0.45\linewidth]{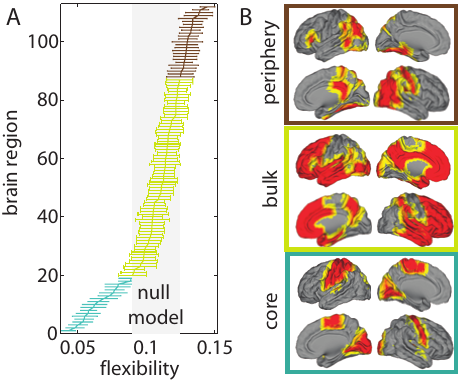}
    \caption{\label{fig:cbpv2} \textbf{Temporal Core-Periphery Organization of the Brain} determined using fMRI signals during the performance of a simple motor learning task. \emph{(A)} The core (cyan), bulk (gold), and periphery (maroon) nodes consist, respectively, of brain regions whose mean flexibility over individuals is less than, equal to, and greater than that expected in a null model (gray shaded region).  We measure flexibility based on the allegiance of nodes to putative functional modules. Error bars indicate the standard error of the mean over individuals. \emph{(B)} The anatomical distribution of regions in the core, bulk, and periphery appears to be spatially contiguous. The core primarily contains sensorimotor and visual processing areas, the periphery primarily contains multimodal association areas, and the bulk contains the remainder of the brain (and is therefore composed predominantly of frontal and temporal cortex).
    }
    \end{center}
    \end{figure}

\paragraph{Regional Variation in Flexibility.} The mean flexibility over participants varied over brain regions.  It ranged from approximately $0.04$ to approximately $0.14$, \rev{which implies that brain regions changed their modular affiliation between 4\% and 14\% of trial blocks on average} (see Fig.~\ref{fig:cbpv2}A). The distribution of flexibility across brain regions is decidedly non-Gaussian: the majority of brain regions have relatively high flexibilities, but there is a left-heavy tail of regions (including a small peak) with low values of flexibility. We characterized the distribution of flexibility over brain regions by calculating the third (skewness) and fourth (kurtosis) central moments. The skewness was $0.50 \pm 0.26 $, and the kurtosis was $3.04 \pm 0.57$. To interpret these findings, we note that a distribution's skewness is a measure of its asymmetry, and the positive values that we observe indicate that the distributions from all participants are skewed to the right. The kurtosis of a distribution is a combined measure of its peakedness \cite{Dodge2003} and its bimodality \cite{Darlington1970}, and it is sometimes construed as a measure of the extent that a distribution is prone to outliers. The kurtosis values that we observe vary between 2.5 and 5, which includes the value of 3 that occurs for a Gaussian distribution.

\paragraph{Defining the Temporal Core and Temporal Periphery.} To determine the significance of a brain region's variation in flexibility, we compared the flexibility of brain regions in the empirical multilayer network to that expected in a nodal null model. We can define a temporal core, bulk, and periphery (see Fig.~\ref{fig:cbpv2}). The core is the set of regions whose flexibility is significantly less than expected in the null model; the periphery is the set of regions whose flexibility is significantly greater than expected in the null model; and the bulk consists of all remaining regions. As discussed in the Text S1, the delineation of the brain into these three groups is robust both to the intensity of training (MIN, MOD, or EXT) and to the duration of training (sessions 1--4). Furthermore, the temporal core, bulk, and periphery tend to form their own communities, although the relationship between core-periphery organization and modular organization appears to be altered by learning (see the Text S1).

%

We show the anatomical locations of the temporal core, temporal bulk, and temporal periphery in Fig.~\ref{fig:cbpv2}. The relatively stiff core is composed of 19 regions located predominantly in primary sensorimotor areas in both left and right hemispheres. Most of the motor-related regions in the core were left-lateralized, which is consistent with the participants' use of their right hand to perform the motor sequence. The more flexible periphery is composed of 25 regions located predominantly in multimodal areas---including inferior parietal, intraparietal sulcus, temporal parietal junction, inferotemporal, fusiform gyrus, and visual association areas. The bulk contains the remaining 68 cortical and subcortical regions---including large swaths of frontal, temporal, and parietal cortices. See Table S3 for a complete list of the affiliation of each brain region to the temporal core, bulk, and periphery. The separation of a temporally stiff core of predominantly unimodal regions that process information from single sensory modality (e.g., vision, audition, etc.) and a flexible periphery of predominantly multimodal cortices that process information from multiple modalities is consistent with existing understanding of the association of multimodal cortex with the binding of different types of information and the performance of a broad range of cognitive functions \cite{Mesulam1998}.

\paragraph{Temporal Core-Periphery Organization and Learning.} One can interpret the anatomical location of the temporally stiff core that consists primarily of unimodal regions and a flexible periphery that consists primarily of multimodal cortices in the context of the known roles of these cortices in similar tasks. The ability to retrieve and rapidly execute complex motor sequences requires extensive practice. These well-learned sequences are known to be generated by ``core'' areas \cite{Bischoff2004,Matsuzaka2007,Picard1997,Kennerley2004}. However, when first learning a sequence, people can use a variety of cognitive strategies that are supported by other brain systems (some of which are located in the periphery to augment performance) \cite{Willingham2002,Destrebecqz2005}. In some cases, these strategies are detrimental to skill retention \cite{Brown2007}. Consequently, we hypothesized that individuals whose core and periphery are distinct---indicating a strong separability of visuomotor and cognitive regions---would learn better than those whose core and periphery were less distinguishable from one another.

To test this hypothesis, we calculated the Spearman rank coefficient $\rho$ between the skewness and kurtosis of the flexibility distribution estimated from the fMRI data of the first scanning session and the learning parameter $\kappa$ estimated over the next 10 days of home training (see Materials and Methods). The kurtosis (in essence) measures the separation between the temporal core and the temporal periphery and is negatively correlated with $\kappa$ (the correlation is $\rho \doteq -0.498$, and the p-value is $p \doteq 0.027$), indicating that individuals with a narrower separation between temporal core and temporal periphery learn better in the subsequent $10$ home training sessions than individuals with a greater separation between temporal core and temporal periphery. Skewness (in essence) measures the presence of---rather than the separation between---the temporal core and the temporal periphery and is also negatively correlated with learning ($\rho \doteq -0.480$ and $p \doteq 0.034$), indicating that individuals whose flexibility was more skewed over brain regions learned better than those whose flexibility over brain regions was less skewed. This finding implies that individuals with ``stronger'' temporal core-periphery structure (i.e., larger values of skewness) learn better than those with ``weaker'' temporal core-periphery structure (i.e., smaller values of skewness).

Importantly, the temporal separation of the data from the scanning session (which we used to estimate brain flexibility) and the home training (which we used to estimate learning) ensures that these correlations are predictive. Together, these results indicate that individuals with a stronger temporal core and temporal periphery but a smooth transition between them seem to learn better than individuals with a weaker temporal core and temporal periphery but a sharper transition between them. These results suggest that successful brain function might depend on a delicate balance between a set of core regions whose allegiance to putative functional modules changes little over time and a set of peripheral regions whose allegiance to putative functional modules is flexible through time (and also on the smoothness of the transition between these two types of regions).


\subsection*{Geometrical Core-Periphery Organization}
\addcontentsline{toc}{subsection}{Geometrical Core-Periphery Organization}

Given our demonstration that there exists a temporal core in dynamic brain networks, it is important to ask what role such core regions might play in individual network layers of the multilayer network \cite{Bassett2011b}. While the roles of nodes in a static network can be studied in multiple ways \cite{Guimera2005,Guimera2007}, we focus on describing the geometrical core-periphery organization---\rev{which can be used to help characterize the organization of edge strengths throughout a network}---to compare it with the temporal core-periphery organization discussed above.  \rev{The geometrical core of a network is composed of a set of regions that are strongly and mutually interconnected. Measures of network centrality can be useful for identifying nodes in a geometrical core because such measures help capture a node's relative importance within a network in terms of its immediate connections, its distance to other nodes in the graph, or its influence on other nodes in the graph \cite{Rombach2012,Opsahl2010}.}

Drawing on studies of social networks \cite{Borgatti1999}, we \rev{examine geometrical core-periphery organization in networks extracted from individual time windows by testing} whether core nodes are densely connected to one another and whether peripheral nodes are sparsely connected to one another. \rev{Rather than proposing a strict separation between a single core and single periphery, we assess the role of a node along a core-periphery spectrum using a centrality measure known as the (geometrical) core score $C$, which was introduced in Ref.~\cite{Rombach2012}. Network nodes with high $C$ values are densely connected to one another, whereas nodes with low $C$ values are sparsely connected to one another.  The method in Ref.~\cite{Rombach2012} uses a two-parameter function to interpolate between core nodes and peripheral nodes. One parameter (which is denoted by $\alpha$) sets the sharpness of the boundary between the geometrical core and the geometrical periphery. Small values of $\alpha$ indicate a fuzzy boundary, whereas large values indicate a sharp transition. The second parameter (which is denoted by $\beta$) sets the size of the geometrical core.  Smaller values of $\beta$ correspond to smaller cores.
 We can quantify the fit of the transition function that defines the set of core scores to the data using a summary diagnostic that is called the $R$-score (see Materials and Methods \rev{for definitions}). Large values of $R$ indicate a good fit and therefore provide confidence that one has uncovered a good estimate of a network's core-periphery organization.}

\begin{figure}
\begin{center}
\includegraphics[width=1\linewidth]{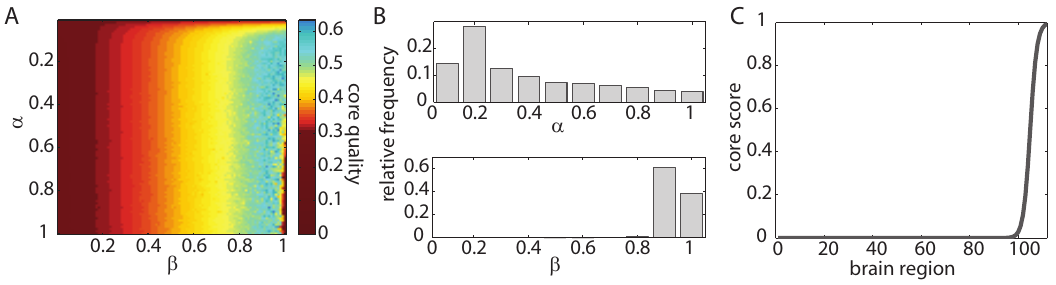}
    \caption[]{\label{fig:puck} \textbf{Geometrical Core-Periphery Organization in Brain Networks}. \emph{(A)} Core quality $R$ (\ref{R}) in the $(\alpha,\beta)$ parameter plane for a typical participant (3), scanning session (1), sequence type (EXT), and experimental block (1). \emph{(B)} Distribution of the $\alpha$ and $\beta$ values that maximize the $R$-score. We compute this distribution over all network layers, participants, scanning sessions, and sequence types. The $\beta$ parameter is much more localized (its standard deviation is 0.05) than the $\alpha$ parameter (its standard deviation is 0.26). \emph{(C)} Mean core shape. We plot the ordered vector of $C$ values.  We have set the values of $\alpha$ and $\beta$ to the mean values of those that maximize the $R$-score for all network layers, participants, scanning sessions, and sequence types.
     }
    \end{center}
\end{figure}

In Fig.~\ref{fig:puck}A, we show a typical $R$-score landscape in the $(\alpha,\beta)$ parameter plane.  This landscape favors a relatively small core and a medium value of the transition-sharpness parameter. To choose sensible values of $\alpha$ and $\beta$ for studying core-periphery organization, we examine the distributions of the relative frequencies of $\alpha$ and $\beta$ values that maximize the $R$-score for each network layer, participant, scanning session, and sequence type (see Fig.~\ref{fig:puck}B). We use the mean values of these distributions ($\alpha \doteq 0.40$ and $\beta \doteq 0.94$) to assign a core score to each node. In Fig.~\ref{fig:puck}C, we show the shape of the ``mean core'' that we obtain using these parameter values. This figure demonstrates that the typical (geometrical) core-periphery organization in the networks under study is a mixture between a discrete core-periphery organization, in which every node is either in the core or in the periphery, and a continuous core-periphery organization, in which there is a continuous spectrum to describe how strongly nodes belong to a core. In these networks (which usually possess a single core), the majority of nodes do not belong to the core, but those nodes that do (roughly $10 \%$ of the nodes) have a continuum level of association strengths with the core.

In some cases, we identified multiple competing cores, which we found by using simulated annealing to explore local maxima of the $R$-score rather than only identifying a global maximum. Because of this stochasticity in the methodology for examining core-periphery organization, we performed computations with the chosen parameter values ($\alpha \doteq  0.40$ and $\beta \doteq  0.94$) 10 times and used the solution with the highest $R$-score out of these 10 iterations for each network layer, subject, scanning session, and sequence type.

\begin{figure}
\begin{center}
\includegraphics[width=.33\linewidth]{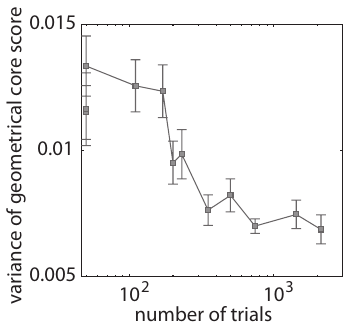}
    \caption[]{\label{fig:scvskeor} \textbf{Geometrical Core Scores Change with Learning}. Variance of the distribution of mean geometrical core scores over brain regions as a function of the number of trials completed after a scanning session. (See Table \ref{experiment0} for the relationship between the number of trials practiced and training duration and intensity.) Error bars indicate the standard error of the mean over participants (where the data point from each participant is the mean geometrical core score over brain regions, scanning sessions, sequence types, and network layers).
       }
    \end{center}
\end{figure}

An interesting question is whether geometrical core-periphery organization remains relatively constant throughout time or whether the organization changes with learning. We observed that regions that have a high geometrical core score in the first scanning session and in EXT blocks were likely to have high geometrical core scores in later scanning sessions and in MOD and MIN blocks. (See the Text S1 for supporting results on the reliability of geometrical core-periphery organization.) In light of this consistency, we calculate a mean geometrical core score for each node by taking the mean over all blocks in a given scanning session (1, 2, 3, and 4) and sequence type (EXT, MOD, and MIN). The variance of the mean geometrical core score over nodes in a network then gives an indication of the separation between the mean core and periphery. As we show in Fig.~\ref{fig:scvskeor}, we find that the variance of the mean geometrical core score over trials decreases as a function of learning. A high variance of the mean core scores over nodes indicates a greater separation between the mean core and periphery as well as a high consistency of the core score of each node over trial blocks. If a node's core score is inconsistent over trial blocks, then the mean core score for each node in the network is expected to be similar and thus one would expect the variance of the core scores over nodes to be small. A low variance in the mean geometrical core score over trial blocks therefore suggests either little separation between the core and periphery or an increased variability in core scores of a given node over trial blocks.



\subsection*{Relationship Between Temporal and Geometrical Core-Periphery Organization}
\addcontentsline{toc}{subsection}{Relationship Between Temporal and Geometrical Core-Periphery Organization}

\begin{figure}
\includegraphics[width=1\linewidth]{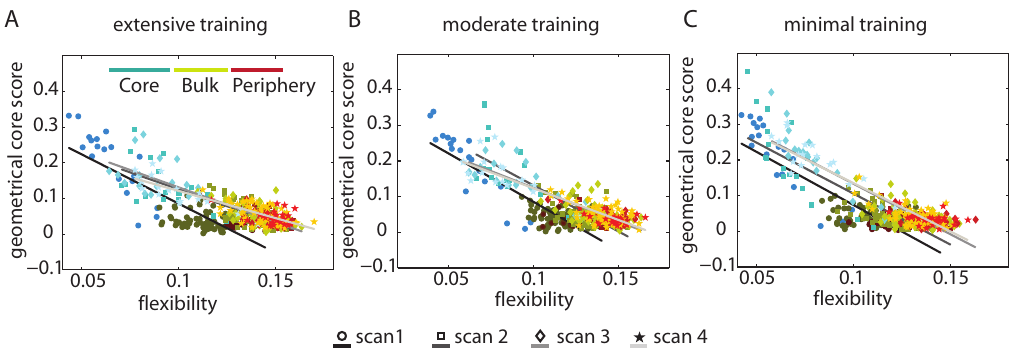}
\begin{center}
    \caption[]{\label{fig:tsc1234eor} \textbf{Relationship Between Temporal and Geometrical Core-Periphery Organizations}. \rev{A strong negative correlation exists between flexibility and the geometrical core score} for networks constructed from blocks of \emph{(A)} extensively, \emph{(B)} moderately, and \emph{(C)} minimally trained sequences on scanning session 1 (day 1; circles), session 2 (after approximately 2 weeks of training; squares), session 3 (after approximately 4 weeks of training; diamonds), and session 4 (after approximately 6 weeks of training; stars).  \rev{This negative correlation indicates that the temporal core-periphery organization is mimicked in the geometrical core-periphery organization and therefore that the core of dynamically stiff regions also exhibits dense connectivity.} We show temporal core nodes in cyan, temporal bulk nodes in gold, and temporal periphery nodes in maroon. The darkness of data points indicates scanning session; darker colors indicate earlier scans, so the darkest colors indicate scan 1 and the lightest  ones indicate scan 4. The grayscale lines indicate the best linear fits; again, darker colors indicate earlier scans, so session 1 is in gray and session 4 is in light gray.
    The Pearson correlation between the flexibility (averaged over 100 multilayer modularity optimizations, 20 participants, and 4 scanning sessions) and the geometrical core score (averaged over 20 participants and 4 scanning sessions) is significant for the EXT ($r\doteq-0.92$, $p\doteq3.4\times 10^{-45}$), MOD ($r\doteq-0.93$, $p\doteq2.2\times 10^{-49}$), and MIN ($r\doteq -0.93$, $p\doteq 4.8\times 10^{-50}$) data.
 }
    \end{center}
\end{figure}

\begin{figure}
\begin{center}
\includegraphics[width=.4\linewidth]{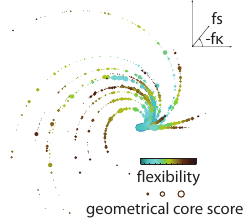}
    \caption{\label{fig:summary} \textbf{Core-Periphery Organization of Brain Dynamics During Learning}. The relationship between temporal and geometrical core-periphery organization and their associations with learning are present in individual subjects. We represent this relationship using spirals in a plane; data points in this plane represent brain regions located at the polar coordinates ($fs$,$-f\kappa$), where $f$ is the flexibility of the region, $s$ is the skewness of flexibility over all regions, and $\kappa$ is the learning parameter (see the Materials and Methods) that describes each individual's relative improvement between sessions. The skewness predicts individual differences in learning; the Spearman rank correlation is $\rho \doteq -0.480$ and $p \doteq 0.034$. Poor learners (straighter spirals) tend to have a low skewness (short spirals), whereas good learners (curvier spirals) tend to have high skewness (long spirals). Color indicates flexibility: blue nodes have lower flexibility, and brown nodes have higher flexibility.
   }
\end{center}
\end{figure}

Given the geometrical core-periphery organization in the individual layers of the multilayer networks and the temporal core-periphery organization in the full multilayer networks, it is important to ask whether brain regions in the temporal core (i.e., regions with low flexibility) are also likely to be in the geometrical core (i.e., whether they exhibit strong connectivity with other core nodes, as represented by a high value of the geometrical core score). In Fig.~\ref{fig:tsc1234eor}, we show scatter plots of the flexibility and core score for the 3 training levels (EXT, MOD, and MIN) and the 4 scanning sessions over the 6-week training period. We find that the temporal core-periphery organization (which is a dynamic measurement) is strongly correlated with the geometrical core score (which is a measure of network geometry and hence of network structure). This indicates that regions with low temporal flexibility tend to be strongly-connected core nodes in (static) network layers. \rev{In Fig.~\ref{fig:tsc1234eor}, we show that the relationship between temporal and geometrical core-periphery organization occurs reliability across training depth, duration, and intensity. In Fig.~\ref{fig:summary}, we show that this relationship can also be identified robustly in data extracted from individual subjects.}



\section*{Discussion}

We have shown how the mesoscale organization of functional brain networks changes over the course of learning. Our results suggest that core-periphery organization is an important and predictive component of cognitive processes that support sequential, goal-directed behavior. \rev{We summarize our findings in Fig.~\ref{fig:summary}, which demonstrates that poor learners tend to have poorer separation between core and periphery (as indicated by straighter, shorter spirals in the figure) and that good learners tend to have greater separation between core and periphery (curvier, longer spirals).} Our findings also demonstrate that during the generation of motor sequences, the brain consists of a temporally stable and densely connected set of core regions complemented by a temporally flexible and sparsely connected set of peripheral regions. This functional tradeoff between a core and periphery might provide a balance between the rigidity necessary to maintain motor function by the core and the adaptivity of the periphery necessary to enable behavioral change as a function of context or strategy.

In the Text S1, we provide supporting results that indicate (i) that our findings are not merely a function of variation in region size and (ii) that they cannot be derived from the underlying block design of the experimental task. We also show in this supplement that (arguably) simpler properties of brain function---such as the regional signal power of brain activity, mean connectivity strength, and parameter estimates from a general linear model---provide less predictive power than core-periphery organization.


\paragraph{Core-Periphery Organization of Human Brain Structure and Function.} The notion of a core-periphery organization is based on the structure (rather than the temporal dynamics) of a network \cite{Rombach2012}.  \rev{Intuitively, a core consists of a set of highly and mutually interconnected set of regions}.  In this paper, we have described what is traditionally called ``core-periphery structure'' using the terminology \emph{geometrical core-periphery organization}.  (It is geometrical rather than topological because the networks are weighted.) This intuitive notion was formalized in social networks by Borgatti and Everett in 1999  \cite{Borgatti1999}.  Available methods to identify and quantify geometrical core-periphery organization in networks include ones based on block models \cite{Borgatti1999}, $k$-core organization \cite{Holme2005}, and aggregation of information about connectivity and short paths through a network  \cite{Silva2008}. Unfortunately, many methods that have been used to study cores and peripheries in networks have binarized networks that are inherently weighted, which requires one to throw away a lot of important information. Even the recently developed weighted extensions of $k$-core decomposition \cite{Hagmann2008} require a discretization of $k$-shells, which have been defined for both binary and weighted networks \cite{Garas2012}. Importantly, $k$-core decomposition is based on a very stringent and specific type of core connectivity, so this measure misses important core-like structures  \cite{Rombach2012,Shanahan2012}. A well-known measure called the ``rich-club coefficient'' (RCC)  \cite{rcc} considers a different but somewhat related question of whether nodes of high degree (defined as $k \geq k_{t}$ some threshold value $k_{t}$) tend to connect to other nodes of high degree.  (The RCC is therefore a form of assortativity.) The RCC has also been extended to weighted networks \cite{Opsahl2008}, but it still requires one to specify a threshold value of richness to enable one to ask whether ``rich'' nodes tend to connect to other ``rich'' nodes.

The aforementioned limitations notwithstanding, several of the measures discussed above have recently been used successfully to identify a structural core of the human brain white-matter tract network, which is characterized not only by a $k$-core with a high value of the degree $k$ (in particular, $k \geq 20$) \cite{Hagmann2008} and rich club \cite{Heuvel2011,Heuvel2012} but also by a knotty center of nodes that have a high geodesic betweenness centrality but not necessarily a high degree \cite{Shanahan2012}. A $k$-core decomposition has also been applied to functional brain imaging data to demonstrate a relationship between network reconfiguration and errors in task performance\cite{Ekman2012}.

A novel approach that is able to overcome many of these conceptual limitations is the geometrical core-score \cite{Rombach2012}, which is an inherently continuous measure, is defined for weighted networks, and can be used to identify regions of a network core without relying solely on their degree or strength (i.e., weighted degree). Moreover, by using this measure, one can produce (i) continuous results, which make it possible to measure whether a brain region is more core-like or periphery-like; (ii) a discrete classification of core versus periphery; or (iii) a finer discrete division (e.g., into 3 or more groups). In addition, this method can identify multiple geometrical cores in a network and rank nodes in terms of how strongly they participate in different possible cores. This sensitivity is particularly helpful for the examination of brain networks for which multiple cores are hypothesized to mediate multimodal integration \cite{Sepulcre2012}. In this paper, we have demonstrated that functional brain networks derived from task-based data acquired during goal-directed brain activity exhibit geometrical core-periphery organization. Moreover, they are specifically characterized by a straightforward core-periphery landscape that includes a relatively small core composed of roughly 10\% or so of the nodes in the network.

In this paper, we have introduced a method and associated definitions to identify a \emph{temporal} core-periphery organization based on changes in a node's module allegiance over time. We have defined the notion of a temporal core as a set of regions that exhibit fewer changes in module allegiance over time than expected in a dynamic-network null model. Neurobiologically, the temporal core contains brain areas that show consistent \revtwo{task-based} mesoscale functional connectivity over the course of an experiment\revtwo{, and it is therefore perhaps unsurprising that their anatomical locations differ from nodes in the $(k\geq20$)-core \cite{Hagmann2008} and RCC \cite{Heuvel2011,Heuvel2012} of the human white matter tract network.} Our approach is inspired by the following idea: although the brain uses the function of a small subset of regions to perform a given task (i.e., some sort of \rev{core}), a set of additional regions that are associated more peripherally with the task might also be activated in a transient manner.  Indeed, several recent studies have highlighted the possibility of a separation between groups of regions that are consistently versus transiently activated during task-related function  \cite{Fornito2012,Dosenbach2006}, and they have demonstrated that correlations between such regions can be altered depending on their activity \cite{Fornito2012,Pasquale2012}.

\rev{Given the very different definitions of the geometrical and temporal cores, it is interesting that nodes in the temporal core are also likely to be present in the geometrical core. Importantly, the notions of temporal and geometrical core are complementary, and they are both intuitive in the context of brain function. A set of regions that is coherently active to perform a task (i.e., is in the geometrical core) must remain online consistently throughout an experiment (i.e., be in the temporal core), whereas a set of regions that might be activated less coherently (i.e., is in the geometrical periphery) can be utilized by separate putative functional modules over time (i.e., it can be in the temporal periphery). This interpretation is consistent with the notion that the anatomical locations of the core and periphery are task-specific. Should brain activity during other tasks also exhibit core-periphery organization, then the core and periphery of these other task networks could consist of a different set of anatomical regions than those observed here. A comparison of dynamic community structure and associated mesoscale organizational properties across brain states elicited by other tasks is outside of the scope of the present study. However, such a study in a controlled sample with similar time-series length and experimental task structure (e.g., trial lengths, block lengths, and rest periods) would likely yield important insights.}


\paragraph{Modular Versus Core-Periphery Organization.} Community structure and core-periphery organization are two types of mesoscale structures, and they can both be present simultaneously in a network  \cite{Rombach2012,Shanahan2012}. Moreover, both modular and core-periphery organization can in principle pertain to different characteristics of or constraints on underlying brain function. In particular, the presence of community structure supports the idea of the brain containing putative functional modules, whereas the presence of a core-periphery organization underscores the fact that different brain regions likely play inherently different roles in information processing. A symbiosis between these two types of organization is highlighted by the findings that we report in this manuscript: the dynamic reconfiguration of putative functional modules can be described parsimoniously by temporal core-periphery organization, demonstrating that one type of mesoscale structure can help to characterize another. Furthermore, the notion that the brain can simultaneously contain functional modules (e.g., the executive network or the default-mode network) and regions that transiently mediate interactions between modules is consistent with recent characterizations of attention and cognitive control processes  \cite{Uddin2011}.


\paragraph{Dynamic Brain Networks.} It is increasingly apparent that functional connectivity in the brain changes over time and that these changes are biologically meaningful. Several recent studies have highlighted the temporal variability \cite{Whitlow2011,Kramer2011,Chu2012,Allen2012} and non-stationarity  \cite{Jones2012} of functional brain network organization, and both of these features are apparent over short time intervals (less than 5 minutes in fMRI; less than 100 s in EEG) \cite{Whitlow2011,Chu2012,Kramer2011}. Although temporal variability in functional connectivity was seen initially as a signature of measurement noise \cite{Jones2012}, recent evidence suggests instead that it might provide an indirect measurement of changing cognitive processes. Thus, it might serve as a diagnostic biomarker of disease \cite{Jones2012,Felix2012}. Moreover, such temporal variability appears to be modulated by exogenous inputs. For example, Barnes et al. \cite{Barnes2009} demonstrated using a continuous acquisition ``rest-task-rest'' design that endogenous brain dynamics do not return to their pre-task state until approximately 18 minutes following task completion. Similar results that consider other tasks have also been reported  \cite{Tambini2010}. More generally, the dynamic nature of brain connectivity is likely linked to spontaneous cortical processing, reflecting a combination of both stable and transient communication pathways \cite{Chu2012,Kramer2011,Doron2012}.


\paragraph{Network Predictions of Future Learning.} In this study, we observed that properties of the temporal organization of functional brain networks (e.g., on day 1 of this experiment) can be used to predict extended motor learning (e.g., on the following 10 days of home training on a discrete sequence-production task). Our findings are consistent with two previous studies that demonstrated a predictive connection between both dynamic \cite{Bassett2011b} and topological \cite{Sheppard2012} network organization and subsequent learning. (Note that we use the term \emph{topological} because Ref.~\cite{Sheppard2012} considered only unweighted networks.) Reference \cite{Bassett2011b} focused on early---rather than extended---learning of a cued sequence-production motor task (rather than a discrete one) and found that network flexibility on the first day of experiments predicted learning on the second day and that flexibility on the second day predicted learning on the third day. Reference \cite{Sheppard2012} investigated participants' success in learning words of an artificial spoken language and found that network properties from individual time windows could be used to predict such success \cite{Sheppard2012}. Together with the present study, these results highlight the potential breadth of the relationship between network organization and learning. The presence of such a relationship has now been identified across multiple tasks, over multiple time scales, and using both dynamic and topological network properties.


\paragraph{Methodological Considerations.}

\rev{Our study has focused on large-scale changes in dynamic community structure that are correlated with learning. Finer-scale investigations that employ alternative parcellation schemes \cite{Zalesky2010,Wang2009,Power2011,Bassett2010c,Wig2011,deReus2013} with greater spatial resolution or alternative neuroimaging techniques such as EEG or MEG \cite{Doron2012} with greater temporal resolution might uncover additional features that would enhance understanding of functional network-based predictors of learning phenomena.}

\rev{Throughout this paper, we have referred to feature similarities (which we estimated using the magnitude squared coherence) between pairs of regional BOLD time series as \emph{functional connectivity} \cite{Friston2011}. As appreciated in prior literature \cite{Horwitz2003,Gavrilescu2008,Jones2010}, the interpretation of functional connectivity must be made with caution. Coherence in the activity recorded at different brain sites does not necessitate that those sites share information with one another to enable cognitive processing, as they could instead indicate that those two sites are activated by the same third party (either another brain region or an external stimulus). In this paper, we do not distinguish between these two possible drivers of strong inter-regional coherence. Future studies could employ multiple estimates of statistical associations in the form of diagnostics \cite{David2004,Smith2011,Ryali2012} and/or models \cite{Gates2012,Watanabe2013} that might uncover other sets of interactions that could predict the observed coherence structure and hence the observed behavior.}

\section*{Materials and Methods}

\subsection*{Experiment and Data Acquisition}
\addcontentsline{toc}{subsection}{Experiment and Data Acquisition}

\subsubsection*{Ethics Statement}

Twenty-two right-handed participants (13 females and 9 males; the mean age was about 24) volunteered with informed consent in accordance with the Institutional Review Board/Human Subjects Committee, University of California, Santa Barbara.


\subsubsection*{Experiment Setup and Procedure}
\addcontentsline{toc}{subsubsection}{Experiment Setup and Procedure}

We excluded two participants from the investigation: one participant failed to complete the experiment, and the other had excessive head motion. Our investigation therefore includes twenty participants, who all had normal/corrected vision and no history of neurological disease or psychiatric disorders. Each of these participants completed a minimum of 30 behavioral training sessions as well as 3 fMRI test sessions and a pre-training fMRI session. Training began immediately following the initial pre-training scan session. Test sessions occurred after every 2-week period of behavioral training, during which at least 10 training sessions were required. The training was done on personal laptop computers using a training module that was installed by the experimenter (N.F.W.). Participants were given instructions for how to run the module, which they were required to do for a minimum of 10 out of 14 days in a 2-week period. Participants were scanned on the first day of the experiment (scan 1), and then a second time approximately 14 days later (scan 2), once again approximately 14 days later (scan 3), and finally 14 days after that (scan 4). Not all participants were scanned exactly every two weeks; see Table S1 for details of the number of days that elapsed between scanning sessions.

\begin{figure}
\begin{center}
\includegraphics[width=0.7\linewidth]{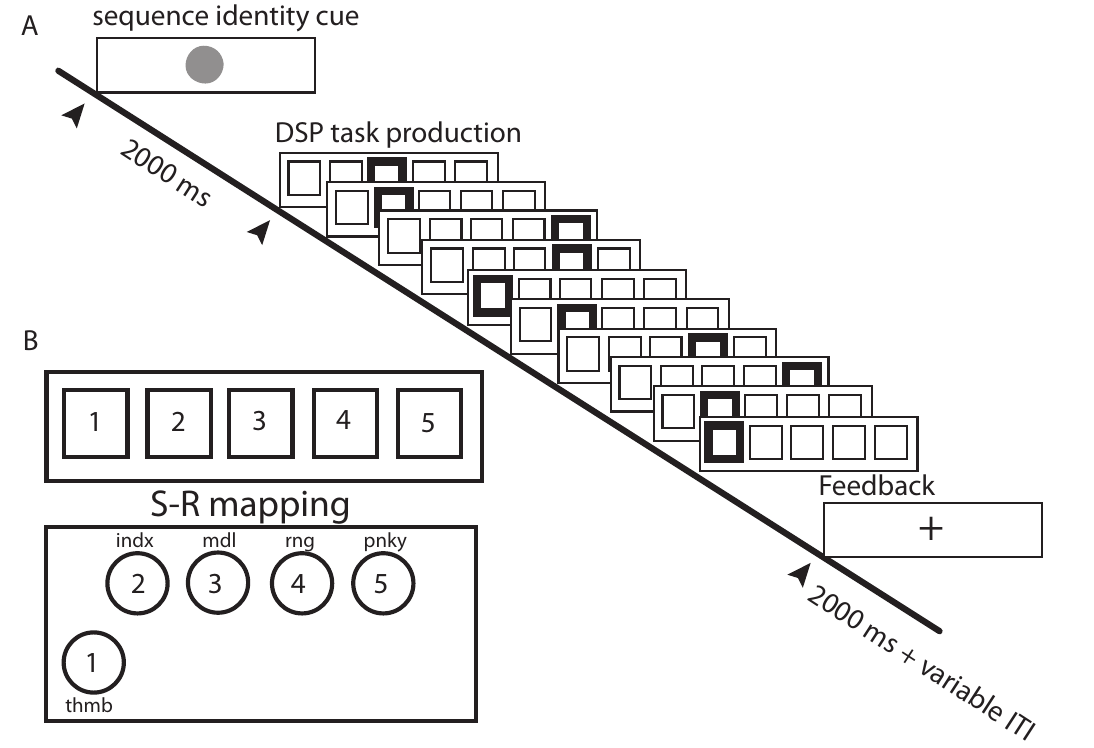}
    \caption[]{\label{fig:dsp} \textbf{Trial Structure and Stimulus-Response (S-R) Mapping}. \emph{(A)} Each trial began with the presentation of a sequence-identity cue that remained on screen for $2$ seconds. Each of the $6$ trained sequences was paired with a unique identity cue. A discrete sequence-production (DSP) event structure was used to guide sequence production. The onset of the initial DSP stimulus (thick square, colored red in the task) served as the imperative to produce the sequence. A correct key press led to the immediate presentation of the next DSP stimulus (and so on) until the $10$-element sequence was correctly executed. Participants received a feedback ``+'' to signal that a sequence was completed and to wait (approximately $0$--$6$ seconds) for the start of the next trial. This waiting period is called the ``inter-trial interval'' (ITI). At any point, if an incorrect key was hit, a participant would receive an error signal (not shown in the figure) and the DSP sequence would pause until the correct response was received. \emph{(B)} There was a direct S-R mapping between a conventional keyboard or an MRI-compatible button box (see the lower left of the figure) and a participant's right hand, so the leftmost DSP stimulus cued the thumb and the rightmost stimulus cued the pinky finger. Note that the button location for the thumb was positioned to the lower left to achieve maximum comfort and ease of motion.
    }
    \end{center}
\end{figure}

We asked participants to practice a set of 10-element sequences that were presented visually using a discrete sequence-production (DSP) task by generating responses to sequentially presented stimuli (see Fig.~\ref{fig:dsp}) using a laptop keyboard with their right hand. Sequences were presented using a horizontal array of 5 square stimuli; the responses were mapped from left to right, such that the thumb corresponded to the leftmost stimulus and the smallest finger corresponded to the rightmost stimulus. A square highlighted in red served as the imperative to respond, and the next square in the sequence was highlighted immediately following each correct key press. If an incorrect key was pressed, the sequence was paused at the error and was restarted upon the generation of the appropriate key press.

%

\begin{figure}
\begin{center}
\includegraphics[width=0.7\linewidth]{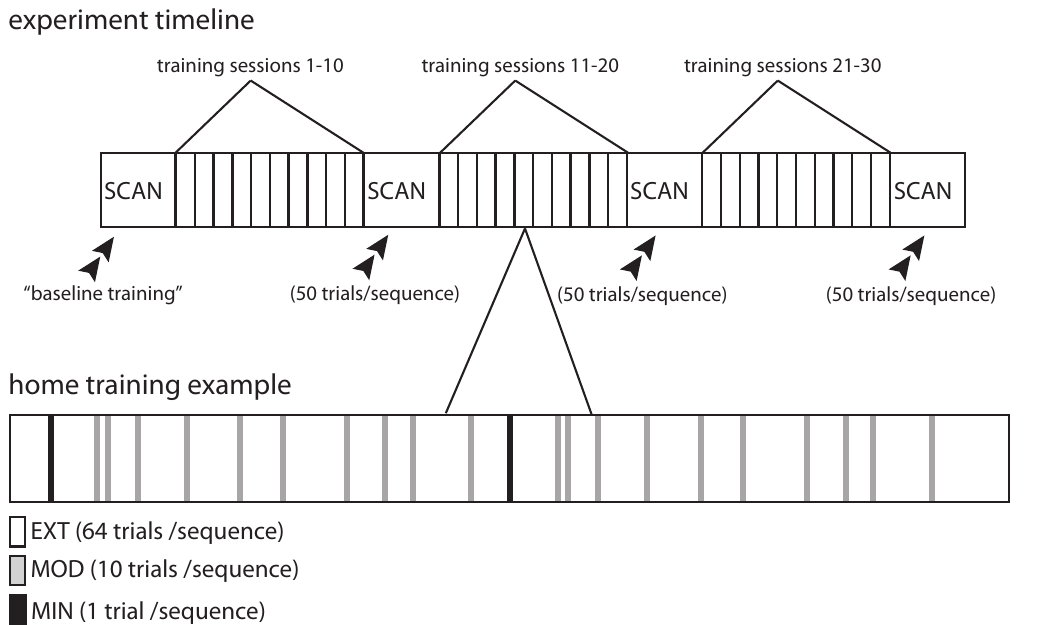}
    \caption[]{\label{fig:exptime} \textbf{Experiment Timeline}. Training sessions in the MRI scanner during the collection of blood-oxygen-level-dependent (BOLD) signals were interleaved with training sessions at home. Participants first practiced the sequences in the MRI scanner during a baseline training session \emph{(top)}. Following every approximately 10 training sessions (see Table S1), participants returned for another scanning session. During each scanning session, a participant practiced each sequence for 50 trials. Participants trained at home between the scanning sessions \emph{(bottom)}. During each home training session, participants practiced the sequences in a random order. (We determined a random order using the Mersenne Twister algorithm of Nishimura and Matsumoto \cite{Matsumoto1998} as implemented in the random number generator {\tt rand.m} of {\sc Matlab} version 7.1). Each EXT sequence was practiced for 64 trials, each MOD sequence was practiced for 10 trials, and each MIN sequence was practiced for 1 trial.
    }
    \end{center}
\end{figure}

Participants had an unlimited amount of time to respond and to complete each trial. All participants trained on the same set of 6 different 10-element sequences, which were presented with 3 different levels of exposure. We organized sequences so that each stimulus location was presented twice and included neither stimulus repetition (e.g., ``11'' could not occur) nor regularities such as trills (e.g., ``121'') or runs (e.g., ``123''). Each training session (see Fig.~\ref{fig:exptime}) included 2 extensively trained sequences (``EXT'') that were each practiced for 64 trials, 2 moderately trained sequences (``MOD'') that were each practiced for 10 trials, and 2 minimally trained sequences (``MIN'') that were each practiced for 1 trial. (See Table S1 for details of the number of trials composed of extensively, moderately, and minimally trained sequences during home training sessions.) Each trial began with the presentation of a sequence-identity cue. The purpose of the identity cue was to inform the participant what sequence they were going to have to type. 
For example, the EXT sequences were preceded by either a cyan (sequence A) or magenta (sequence B) circle.  Participants saw additional identity cues for the MOD sequences (red or green triangles) and for the MIN sequences (orange or white stars, each of which was outlined in black). No participant reported any difficulty viewing the different identity cues. Feedback was presented after every block of 10 trials; this feedback detailed the number of error-free sequences that the participant produced and the mean time it took to complete an error-free sequence.


Each fMRI test session was completed after approximately 10 home training sessions (see Table S1 for details of the number of home practice sessions between scanning sessions), and each participant participated in 3 test sessions. In addition, each participant had a pre-training scan session that was identical to the other test scan sessions immediately prior to the start of training (see Fig.~\ref{fig:exptime}). To familiarize participants with the task, we gave a brief introduction prior to the onset of the pre-training session. We showed the participants the mapping between the fingers and the DSP stimuli, and we explained the significance of the sequence-identity cues.

To help ease the transition between each participant's training environment and that of the scanner, padding was placed under his/her knees to maximize comfort. Participants made responses using a fiber-optic response box that was designed with a similar configuration of buttons as those found on the typical laptop used during training. See the lower left of Fig.~\ref{fig:dsp} for a sketch of the button box used in the experiments. For instance, the center-to-center spacing between the buttons on the top row was 20 mm (compared to 20 mm from ``G'' to ``H'' on a recent MacBook Pro), and the spacing between the top row and lower left ``thumb'' button was 32 mm (compared to 37 mm from ``G'' to the spacebar on a MacBook Pro). The response box was supported using a board whose position could be adjusted to accommodate a participant's reach and hand size. Additional padding was placed under the right forearm to minimize muscle strain when a participant performed the task. Head motion was minimized by inserting padded wedges between the participant and the head coil of the MRI scanner. The number of sequence trials performed during each scanning session was the same for all participants, except for two abbreviated sessions that resulted from technical problems. In each case that scanning was cut short, participants completed 4 out of the 5 scan runs for a given session. We included data from these abbreviated sessions in this study.

Participants were tested inside of the scanner with the same DSP task and the same 6 sequences that they performed during training. Participants were given an unlimited time to complete trials, though they were instructed to respond quickly but also to maintain accuracy. Trial completion was signified by the visual presentation of a fixation mark ``+'', which remained on the screen until the onset of the next sequence-identity cue. To acquire a sufficient number of events for each exposure type, all sequences were presented with the same frequency. Identical to training, trials were organized into blocks of 10 followed by performance feedback. Each block contained trials belonging to a single exposure type and included 5 trials for each sequence. Trials were separated by an inter-trial interval (ITI) that lasted between 0 and 6 seconds (not including any time remaining from the previous trial). Scan epochs contained 60 trials (i.e., 6 blocks) and consisted of 20 trials for each exposure type. Each test session contained 5 scan epochs, yielding a total of 300 trials and a variable number of brain scans depending on how quickly the task was performed. See Table S2 for details of the number of scans in each experimental block.


\subsubsection*{Behavioral Apparatus}
\addcontentsline{toc}{subsubsection}{Behavioral Apparatus}

Stimulus presentation was controlled during training using a participant's laptop computer, which was running Octave 3.2.4 (an open-source program that is very similar to {\sc Matlab}) in conjunction with PsychtoolBox Version 3. We controlled test sessions using a laptop computer running {\sc Matlab} version 7.1 (Mathworks, Natick, MA). We collected key-press responses and response times using a custom fiber-optic button box and transducer connected via a serial port (button box: HHSC-$1\times4$-L; transducer: fORP932; Current Designs, Philadelphia, PA).


\subsubsection*{Behavioral Estimates of Learning}
\addcontentsline{toc}{subsubsection}{Behavioral Estimates of Learning}

Our goal was to study the relationship between brain organization and learning. To ensure independence of these two variables, we extracted brain network structure during the 4 scanning sessions, and we extracted behavioral estimates of learning in home training sessions $1$--$10$ (approximately between days 1 and 14; see Table S1), which took place before scanning session $2$.

For each sequence, we defined the movement time (MT) as the difference between the time of the first button press and the time of the last button press during a single sequence. For the set of sequences of a single type (i.e., sequence 1, 2, 3, 4, 5, or 6), we estimated the learning rate by fitting an exponential function (plus a constant) to the MT data \cite{Schmidt2005,Rosenbaum2010} using a robust outlier correction in {\sc Matlab} (using the function {\tt fit.m} in the Curve Fitting Toolbox with option ``Robust'' and type ``Lar''):
\begin{equation}\label{expon}
	MT = D_1 e^{t/\kappa} + D_2\,,
\end{equation}
where $t$ is time, $\kappa$ is the exponential dropoff parameter (which we call the ``learning parameter'') used to describe the early (and fast) rate of improvement, and $D_1$ and $D_2$ are real and positive constants. The sum $D_1+D_2$ is an estimate of the starting speed of a given participant prior to training, and the parameter $D_2$ is an estimate of the fastest speed attainable by that participant after extended training. A negative value of $\kappa$ indicates a decrease in MT, which is thought to indicate that learning is occurring \cite{Yarrow2009,Dayan2011}. This decrease in MT has been used to quantify learning for several decades \cite{Snoddy1926,Crossman1959}. Several functional forms have been suggested for the fit of MT \cite{Newell1981,Heathcote2000}, and the exponential (plus constant) is viewed as the most statistically robust choice \cite{Heathcote2000}. Additionally, the fitting approach that we used has the advantage of estimating the rate of learning independent of initial performance or performance ceiling.


\subsection*{Functional MRI (fMRI) Imaging}
\addcontentsline{toc}{subsection}{Function MRI (fMRI) Imaging}


\subsubsection*{Imaging Procedures}
\addcontentsline{toc}{subsubsection}{Imaging Procedures}

We acquired
fMRI
signals using a 3.0 T Siemens Trio with a 12-channel phased-array head coil. For each scan epoch, we used a single-shot echo planar imaging sequence that is sensitive to
BOLD
contrast to acquire 37 slices per repetition time (TR of 2000 ms, 3 mm thickness, 0.5 mm gap) with an echo time (TE) of 30 ms, a flip angle of 90 degrees, a field of view (FOV) of 192 mm, and a $64 \times 64$ acquisition matrix. Before the collection of the first functional epoch, we acquired a high-resolution T1-weighted sagittal sequence image of the whole brain (TR of 15.0 ms, TE of 4.2 ms, flip angle of 9 degrees, 3D acquisition, FOV of 256 mm, slice thickness of 0.89 mm, and $256 \times 256$ acquisition matrix).


\subsubsection*{fMRI Data Preprocessing}
\addcontentsline{toc}{subsubsection}{fMRI Data Preprocessing}

We processed  and analyzed functional imaging data using Statistical Parametric Mapping (SPM8, Wellcome Trust Center for Neuroimaging and University College London, UK). We first realigned raw functional data, then coregistered it to the native T1 (normalized to the MNI-152 template with a re-sliced resolution of $3 \times 3 \times 3$ mm), and finally smoothed it using an isotropic Gaussian kernel of 8 mm full-width at half-maximum. To control for potential fluctuations in signal intensity across the scanning sessions, we normalized global intensity across all functional volumes.


\subsection*{Network Construction}
\addcontentsline{toc}{subsection}{Network Construction}


\subsubsection*{Partitioning the Brain into Regions of Interest}
\addcontentsline{toc}{subsubsection}{Partitioning the Brain into Regions of Interest}

Brain function is characterized by spatial specificity: different portions of the cortex emit different, task-dependent activity patterns. To study regional specificity of the functional time series and putative interactions between brain areas, it is common to apply a standardized atlas to raw fMRI data \cite{Bassett2006b,Bassett2009b,Bullmore2009}. \rev{The choice of atlas or parcellation scheme is the topic of several recent studies in structural \cite{Bassett2010c,Zalesky2010}, resting-state \cite{Wang2009}, and task-based \cite{Power2011} network architecture. The question of the most appropriate delineation of the brain into nodes of a network is an open one and is guided by the particular scientific question at hand \cite{Bullmore2011,Wig2011}.}

\rev{Consistent with previous studies of task-based functional connectivity during learning \cite{Bassett2011b,Bassett2012b,Mantzaris2013}}, we parcellated the brain into 112 identifiable cortical and subcortical regions using the structural Harvard-Oxford (HO) atlas (see Table S3) installed with the FMRIB (Oxford Centre for Functional Magnetic Resonance Imaging of the Brain) Software Library (FSL; Version 4.1.1) \cite{Smith2004,Woolrich2009}. For each individual participant and each of the 112 regions, we determined the regional mean BOLD time series by separately averaging across all of the voxels in that region.

Within each HO-atlas region, we constrained voxel selection to voxels that are located within an individual participant's gray matter. To do this, we first segmented each individual participant's T1 into white and gray matter volumes using the DARTEL toolbox supplied with SPM8.  We then restricted the gray-matter voxels to those with an intensity of 0.3 or more (the maximum intensity was 1.0). Note that units are based on an arbitrary scale. We then spatially normalized the participant T1 and corresponding gray matter volume to the MNI-152 template---using the standard SPM 12-parameter affine registration from the native images to the MNI-152 template image---and resampled to 3 mm isotropic voxels. We then restricted the voxels for each HO region by using the program fslmaths \cite{Smith2004,Woolrich2009} to include only voxels that are in the individual participant's gray-matter template.


\subsubsection*{Wavelet Decomposition}
\addcontentsline{toc}{subsubsection}{Wavelet Decomposition}

Brain function is also characterized by frequency specificity. Different cognitive and physiological functions are associated with different frequency bands, and this can be investigated using wavelets. Wavelet decompositions of fMRI time series have been applied extensively in both resting-state and task-based conditions \cite{Bullmore2003,Bullmore2004}. In both cases, they provide sensitivity for the detection of small signal changes in non-stationary time series with noisy backgrounds \cite{Brammer1998}. In particular, the maximum-overlap discrete wavelet transform (MODWT) has been used extensively in connectivity investigations of fMRI \cite{Achard2006,Bassett2006a,Achard2007,Achard2008,Bassett2009,Lynall2010}.  Accordingly, we used MODWT to decompose each regional time series into wavelet scales corresponding to specific frequency bands \cite{Percival2000}.

We were interested in quantifying high-frequency components of an fMRI signal, correlations between which might be indicative of cooperative temporal dynamics of brain activity during a task. Because our sampling frequency was 2 seconds (1 TR = 2 sec), wavelet scale one provides information on the frequency band 0.125--0.25 Hz and wavelet scale two provides information on the frequency band 0.06--0.125 Hz. Previous work has indicated that functional associations between low-frequency components of the fMRI signal (0--0.15 Hz) can be attributed to task-related functional connectivity, whereas associations between high-frequency components (0.2--0.4 Hz) cannot \cite{Sun2004}. This frequency specificity of task-relevant functional connectivity is likely due at least in part to the hemodynamic response function, which might act as a noninvertible band-pass filter on underlying neural activity \cite{Sun2004}. Consistent with our previous work \cite{Bassett2011b}, we examined wavelet scale two, which is thought to be particularly sensitive to dynamic changes in task-related functional brain architecture.


\subsubsection*{Construction of Dynamic Networks}
\addcontentsline{toc}{subsubsection}{Construction of Dynamic Networks}

For each of the 112 brain regions, we extracted the wavelet coefficients of the mean time series in temporal windows given by trial blocks (of approximately 60 TRs; see Table S2). The leftmost temporal boundary of each window was equal to the first TR of an experimental trial block, and the rightmost boundary was equal to the last TR in the same block. We thereby extracted block-specific data sets from the EXT, MOD, and MIN sequences (with 6--10 blocks of each sequence type; see Table S2 for details of the number of blocks of each sequence type) for each of the 20 participants participating in the experiment and for each of the 4 scanning sessions.

For each block-specific data set, we constructed an $N\times N$ adjacency matrix ${\bf W}$ representing the complete set of pairwise functional connections present in the brain during that window in a given participant and for a given scan. Note that $N = 112$ is the number of brain regions in the full brain atlas (see the earlier section on ``Partitioning the Brain into Regions of Interest'' for further details). To quantify the weight $W_{ij}$ of functional connectivity between regions labeled $i$ and $j$, we used the magnitude squared spectral coherence as a measure of nonlinear functional association between any two wavelet coefficient time series (consistent with our previous study \cite{Bassett2011b}). In using the coherence, which has been demonstrated to be useful in the context of fMRI neuroimaging data \cite{Sun2004}, we were able to measure frequency-specific linear relationships between time series.

To examine changes in functional brain network architecture during learning, we constructed multilayer networks by considering the set of $L$ adjacency matrices constructed from consecutive blocks of a given sequence type (EXT, MOD, or MIN) in a given participant and scanning session. We combined the matrices in each set separately to form a rank-3 adjacency tensor ${\bf A}$ per sequence type, participant, and scan.  Such a tensor can be used to represent a time-dependent network \cite{Mucha2010,Bassett2011b}. In the following sections, we describe a variety of diagnostics that can be used to characterize such multilayer structures.


\subsection*{Network Examination}
\addcontentsline{toc}{subsection}{Network Examination}

\subsubsection*{Dynamic Community Detection}
\addcontentsline{toc}{subsubsection}{Dynamic Community Detection}

Community detection \cite{Porter2009,Fortunato2010} can be used to identify putative functional modules (i.e., sets of brain regions that exhibit similar trajectories through time). One such technique is based on the optimization of the modularity quality function \cite{NG2004,Newman2006,Newman2006b}.  This allows one to identify groups that consist of nodes that have stronger connections among themselves than they do to nodes in other groups \cite{Porter2009}. Recently, the modularity quality function has been generalized so that one can consider time-dependent or multiplex networks using \emph{multilayer modularity} \cite{Mucha2010}
\begin{equation} \label{eq:Qml}
    	Q = \frac{1}{2\mu}\sum_{ijlr}\left\{\left(A_{ijl}-\gamma_l P_{ijl} \right)\delta_{lr} + \delta_{ij}\omega_{jlr}\right\} \delta(g_{il},g_{jr})\,,
\end{equation}
where the adjacency matrix of layer $l$ has components $A_{ijl}$, the element
$P_{ijl}$ gives the components of the corresponding matrix for a null model, $\gamma_l$ is the structural resolution parameter of layer $l$, the quantity $g_{il}$ gives the community (i.e., ``module'') assignment of node $i$ in layer $l$, the quantity $g_{jr}$ gives the community assignment of node $j$ in layer $r$, the parameter $\omega_{jlr}$ is the connection strength---i.e., ``interlayer coupling parameter'', which gives an element of a tensor $\mathbf{\omega}$ that constitutes a set of \emph{temporal resolution parameters} if one is using the adjacency tensor ${\bf A}$ to represent a time-dependent network---between
node $j$ in layer $r$ and node $j$ in layer $l$, the total edge weight in the network is $\mu=\frac{1}{2}\sum_{jr} \kappa_{jr}$, the strength of node $j$ in layer $l$ is $\kappa_{jl}=k_{jl}+c_{jl}$, the intra-layer strength of node $j$ in layer $l$ is $k_{jl}$, and the inter-layer strength of node $j$ in
layer $l$ is $c_{jl} = \sum_r \omega_{jlr}$. We employ the Newman-Girvan null model within each layer by using
\begin{equation}
	P_{ijl}= \frac{k_{il} k_{jl}}{2 m_{l}}\,,
\end{equation}
where $m_{l}=\frac{1}{2} \sum_{ij} A_{ijl}$ is the total edge weight in layer $l$. We let $\omega_{jlr} \equiv \omega = \mbox{constant}$ for neighboring layers (i.e., when $| l - r | = 1$) and $\omega_{jlr} = 0$ otherwise.  We also let $\gamma_l = \gamma = \mbox{constant}$.  In the main text, we report results for $\omega = 1$ and $\gamma = 1$, and we evaluate the dependence of our results on $\gamma$ and $\omega$ in the Text S1.

Optimization of multilayer modularity (\ref{eq:Qml}) yields a partition of the brain regions into communities for each time window. To measure changes in the composition of communities across time (i.e., across experimental blocks), we defined the \emph{flexibility} $f_{i}$ of a node $i$ to be the number of times that a node changed community assignment throughout the set of time windows represented by the multilayer network \cite{Bassett2011b} normalized by the total number of changes that were possible (i.e., by the number of contiguous pairs of layers in the multilayer framework, which in this study ranged from $4$ to $10$; see Table S2). We then defined the flexibility of the entire network as the mean flexibility over all nodes in the network: $F = \frac{1}{N} \sum_{i=1}^{N} f_{i}$. To examine the relationship between brain network flexibility and learning, we confined ourselves to the two EXT (i.e., extensively trained) sequences, in which learning occurs more rapidly than in MOD and MIN sequences. We therefore estimated flexibility from the multilayer networks constructed from blocks of the two EXT sequences in the first scanning session.


\subsubsection*{Identification of Temporal Core, Bulk, and Periphery}
\addcontentsline{toc}{subsubsection}{Identification of Temporal Core, Bulk, and Periphery}

We find that different brain regions have different flexibilities. To determine whether a particular brain region is more or less flexible than expected, we constructed a \emph{nodal null model}, which can be used to probe the individual roles of nodes in a network \cite{Bassett2011b,Bassett2012b}. (Note that alternative null models can be used to probe other aspects of the temporal or geometrical structure in a multilayer network \cite{Bassett2011b,Bassett2012b}.) We rewired the ends of the multilayer network's inter-layer edges (which connect nodes in one layer to nodes in another) uniformly at random. After applying the associated permutation, an inter-layer edge can, for example, connect node $i$ in layer $t$ with node $j \neq i$ in layer $t+1$ rather than being constrained to connect each node $i$ in layer $t$ with itself in layer $t+1$.

We considered 100 different rewirings to construct an ensemble of 100 nodal null-model multilayer networks for each single multilayer network constructed from the brain data. We then estimated the flexibility of each node in each nodal null-model network. We created a distribution of expected mean nodal flexibility values by averaging flexibility over 100 rewirings and the 20 participants. We similarly estimated the mean nodal flexibility of the brain data by averaging flexibility over the 20 participants and 100 optimizations.  (We optimized multilayer modularity using a Louvain-like locally greedy method \cite{Jutla2011,Blondel2008}.  This procedure is not deterministic, so different runs of the optimization procedure can yield slightly different partitions of a network.) We considered a region to be a part of the temporal ``core'' if its mean nodal flexibility was below the 2.5\% confidence bound of the null-model distribution, and we considered a region to be a part of the temporal ``periphery'' if its mean nodal flexibility was above the 97.5\% confidence bound of the null-model distribution. Finally, we considered a region to be a part of the temporal ``bulk'' if its mean nodal flexibility was between the 2.5\% and 97.5\% confidence bounds of the null-model distribution.


\subsubsection*{Geometrical Core-Periphery Organization}
\addcontentsline{toc}{subsubsection}{Geometrical Core-Periphery Organization}

To estimate the geometrical core-periphery organization of the (static) networks defined by each experimental block (i.e., for each layer of a multilayer network), we used the method that was recently proposed in Ref.~\cite{Rombach2012}. This method results in a ``core score'' (which constitutes a centrality measure) for each node that indicates where it lies on a continuous spectrum of roles between core and periphery. This method has numerous advantages over previous formulations used to study core-periphery organization.  In particular, it can identify multiple geometrical cores in a network, which makes it possible to take multiple cores into account and in turn enables one to construct a detailed description of geometrical core-periphery organization by ranking the nodes in terms of how strongly they participate in different possible cores.  Importantly, the continuous nature of the measure removes the need to use an artificial dichotomy of being strictly a core node versus strictly a peripheral node.

In applying method, we consider a vector $C$ with non-negative values, and we let $C_{ij}=C_i \times C_j$, where $i$ and $j$ are two nodes in an $N$-node network. We then seek a core vector $C$ that satisfies the normalization condition
\begin{equation*}
	\sum_{i,j}C_iC_j = 1
\end{equation*}	
and is a permutation of the vector $C^*$ whose components specify the \emph{local (geometrical) core values}
\begin{equation}
	C^*_m=\frac{1}{1+\exp\left\{-(m-N\beta) \times \tan(\pi \alpha /2)\right\}}\,, \quad m \in \{1, \ldots, N\}\,.
\end{equation}	
We seek a permutation that maximizes the \emph{core quality}
\begin{equation}\label{R}
	R = \sum_{i,j}A_{ij}C_iC_j\,.
\end{equation}
This method to compute core-periphery organization has two parameters: $\alpha \in [0,1]$ and $\beta \in [0,1]$. The parameter $\alpha$ sets the sharpness of the boundary between the geometrical core and the geometrical periphery. The value $\alpha = 0$ yields the fuzziest boundary, and $\alpha = 1$ gives the sharpest transition (i.e., a binary transition): as $\alpha$ varies from $0$ to $1$, the maximum slope of $C^*$ varies from $0$ to $+\infty$.  The parameter $\beta$ sets the size of the geometrical core: as $\beta$ varies from $0$ to $1$, the number of nodes included in the core varies from $N$ to $0$. One now has the choice of either taking into account the local core scores of a node for a set of $(\alpha,\beta)$ coordinates sampled from $[0,1] \times [0,1]$ (where one weighs each choice by its corresponding value of $R$) or one can take into account only the score for particular choices of $(\alpha,\beta)$.


\subsection*{Statistics and Software}
\addcontentsline{toc}{subsection}{Statistics and Software}

We performed all data analysis and statistical tests in {\sc Matlab}. We performed the dynamic community detection procedure using freely available {\sc Matlab} code \cite{Jutla2011} that optimizes multilayer modularity using a Louvain-like locally greedy algorithm \cite{Blondel2008}.

\section*{Acknowledgements}

We thank Jean M. Carlson, Matthew Cieslak, John Doyle, Daniel Greenfield, and Megan T. Valentine for helpful discussions; John Bushnell for technical support; and James Fowler, Sang Hoon Lee, and Melissa Lever for comments on earlier versions of the manuscript.


\bibliography{bibfile4}


\section*{Figure and Table Legends}

\begin{figure}
\begin{center}
    \caption[]{\label{fig:cbp1eor} \textbf{(Supplementary Figure S1) Reliability of Temporal Core-Periphery Structure.} Temporal core (cyan), bulk (gold), and periphery (maroon) of dynamic networks determined based on the flexibility of trial blocks in which participants practiced sequences that would eventually be extensively trained. \emph{(A)} Flexibility of the temporal core, bulk, and periphery averaged over the 100 multilayer modularity optimizations and 20 participants for blocks composed of extensively trained (EXT; light circles), moderately trained (MOD; squares), and minimally trained (MIN; dark diamonds) sequences. The darkness of data points indicates scanning session; darker colors indicate earlier scans, so the darkest colors indicate scan 1 and the lightest ones indicate scan 4. \emph{(B)} The coefficient of variation of flexibility calculated over the 100 optimizations and 3 sequence types for all brain regions. Error bars indicate the standard error of the mean CV over participants. Both panels use data from scanning session 1 on day 1 of the experiment (which is prior to home training).
    }
    \end{center}
\end{figure}

\begin{figure}
\begin{center}
    \caption[]{\label{fig:cbp234eor} \textbf{(Supplementary Figure S2) Temporal Core-Periphery Organization Over 42 Days.} Temporal core (cyan), bulk (gold), and periphery (maroon) of dynamic networks defined by trial blocks in which participants practiced sequences that would eventually be \emph{(A)} extensively trained, \emph{(B)} moderately trained, and \emph{(C)} minimally trained for data from scanning sessions 2 (after approximately 2 weeks of training; circles), 3 (after approximately 4 weeks of training; squares), and 4 (after approximately 6 weeks of training; diamonds). The darkness of data points indicates scanning session; darker colors indicate earlier scans, so the darkest colors indicate scan 1 and the lightest ones indicate scan 4.}
    \end{center}
\end{figure}

\begin{figure}
\begin{center}
    \caption[]{\label{fig:sceorreg} \textbf{(Supplementary Figure S3) Geometrical Core-Periphery Organization Over 42 Days.} Geometrical core scores for each brain region defined by the trial blocks in which participants practiced sequences that would eventually be \emph{(A)} extensively trained, \emph{(B)} moderately trained, and \emph{(C)} minimally trained for data from scanning sessions 1 (day 1; black circles), 2 (after approximately 2 weeks of training; dark gray squares), 3 (after approximately 4 weeks of training; gray diamonds), and 4 (after approximately 6 weeks of training; light gray stars). We have averaged the geometrical core scores over blocks and over 20 participants. The order of brain regions is identical for all 3 panels (\emph{A-C}), and we chose this order by ranking regions from high to low geometrical core scores from the EXT blocks on scanning session 1 (on day 1 of the experiment).}
    \end{center}
\end{figure}

\begin{figure}
\begin{center}
    \caption[]{\label{fig:modcbp} \textbf{(Supplementary Figure S4) Relationship Between Temporal Core-Periphery Organization and Community Structure.} \emph{(A)} Mean-coherence matrix over all EXT blocks from all participants on scanning day 1. The colored bars above the matrix indicate the 3 communities that we identified from the representative partition. Mean partition similarity z-score $z_{i}$ over all participants for blocks of \emph{(B)} extensively, \emph{(C)} moderately, and \emph{(D)} minimally trained sequences for all 4 scanning sessions over the approximately 6 weeks of training. The horizontal gray lines in panels \emph{(B-D)} indicate the $z_{i}$ value that corresponds to a right-tailed p-value of $0.05$.
    }
    \end{center}
\end{figure}

\begin{figure}
\begin{center}
    \caption[]{\label{fig:regsiz} \textbf{(Supplementary Figure S5) Region Size is Uncorrelated with Flexibility}. \emph{(A)} Scatter plot of the size of the brain region in voxels (averaged over participants) versus the flexibility of the EXT multilayer networks, which we averaged over the 100 multilayer modularity optimizations and the 20 participants. Data points indicate brain regions. The line indicates the best linear fit.  Its Pearson correlation coefficient is $r \doteq -0.009$, and the associated p-value is $p \doteq 0.92$. \emph{(B)} Box plot over the 20 participants of the squared Pearson correlation coefficient $r^2$ between the participant-specific region size in voxels and the participant-specific flexibility averaged over the 100 multilayer modularity optimizations.}
    \end{center}
\end{figure}

\begin{figure}
\begin{center}
    \caption[]{\label{fig:glmcbp} \textbf{(Supplementary Figure S6) Temporal Core-Periphery Organization and Task-Related Activations.} Mean GLM parameter estimates for the temporal core (cyan; circles), bulk (gold; squares), and periphery (maroon; diamonds) of dynamic networks defined by the trial blocks in which participants practiced sequences that would eventually be \emph{(A)} extensively trained, \emph{(B)} moderately trained, and \emph{(C)} minimally trained for data from scanning sessions 1 (first day of training),  2 (after approximately 2 weeks of training), 3 (after approximately 4 weeks of training), and 4 (after approximately 6 weeks of training).
    }
    \end{center}
\end{figure}

\begin{figure}
\begin{center}
    \caption[]{\label{fig:gamma} \textbf{(Supplementary Figure S7) Effect of Structural Resolution Parameter.} \emph{(A,B)} Number of communities and \emph{(C,D)} number of regions in the temporal core (cyan; circles), temporal bulk (gold; squares), and temporal periphery (maroon; diamonds) as a function of the structural resolution parameter $\gamma$, where we considered \emph{(A,C)} $\gamma \in [0.2,5]$ in increments of $\Delta \gamma = 0.2$ and \emph{(B,D)} $\gamma \in [0.8,1.8]$ in increments of $\Delta \gamma = 0.01$. We averaged the values in panels \emph{(A)} and \emph{(B)} over 100 multilayer modularity optimizations and over the 20 participants.
    }
    \end{center}
\end{figure}

\begin{figure}
\begin{center}
    \caption[]{\label{fig:omega} \textbf{(Supplementary Figure S8) Effect of Temporal Resolution Parameter.} \emph{(A)} Number of communities averaged over 100 multilayer modularity optimizations and over 20 participants as a function of the temporal resolution parameter $\omega$. \emph{(B)} Number of regions that we identified as part of the temporal core (cyan; circles), temporal bulk (gold; squares), and temporal periphery (maroon; diamonds) as we vary $\omega$ from $0.1$ to $2$ in increments of $\Delta \omega = 0.1$.
    }
    \end{center}
\end{figure}

\clearpage
\newpage
\newpage

~\\
~\\
\textbf{(Main Text) Table 1: Relationship Between Training Duration, Intensity, and Depth}. We report the number of trials (i.e., ``depth'') of each sequence type (i.e., ``intensity'') completed after each scanning session (i.e., ``duration'') averaged over the 20 participants.
~\\
~\\

\textbf{(Supplementary Table S1): Experimental Details for Behavioral Data Acquired Between Scanning Sessions}. We give the minimum, mean, maximum, and standard error of the mean over participants for the following variables: the number of days between scanning sessions; the number of practice sessions performed at home between scanning sessions; and the number of trials composed of extensively, moderately, and minimally trained sequences during home practice between scanning sessions.

~\\
~\\

\textbf{(Supplementary Table S2): Experimental Details for Brain Imaging Data Acquired During Scanning Sessions}. In the top three rows, we give the mean, minimum, maximum, and standard error over participants for the number of blocks composed of extensively, moderately, and minimally trained sequences during scanning sessions.  In the bottom three rows, we give (in TRs) the mean, minimum, maximum, and standard error of the length over blocks composed of extensively, moderately, and minimally trained sequences during scanning sessions.

~\\
~\\

\textbf{(Supplementary Table S3): Brain regions in the Harvard-Oxford (HO) Cortical and Subcortical Parcellation Scheme provided by FSL \cite{Smith2004,Woolrich2009}} and their affiliation to the temporal core (C; cyan), bulk (B; gold), and periphery (P; maroon) for both left (L) and right (R) hemispheres.

\begin{center}
\begin{table*}[ht]
{\normalsize
\hfill{}
\begin{tabular}{|c||c|c|c|c|}
\hline
~ &                     Session 1    & Session 2   & Session 3   & Session 4 \\
\hline
\hline
MIN Sequences        & 50     & 110         & 170        & 230          \\ \hline
MOD Sequences        & 50     & 200         & 350        & 500          \\ \hline
EXT Sequences        & 50     & 740         & 1430       & 2120         \\
\hline
\hline
\end{tabular}}
\hfill{}
\caption{\textbf{Relationship Between Training Duration, Intensity, and Depth}. We report the number of trials (i.e., ``depth'') of each sequence type (i.e., ``intensity'') completed after each scanning session (i.e., ``duration'') averaged over the 20 participants.
\label{experiment0}
}
\end{table*}
\end{center}



\end{document}



\title{Supplementary Material for ``Core-Periphery Organization of Human Brain Dynamics"}

\author{Danielle S. Bassett$^{1,2,*}$, Nicholas F.
Wymbs$^{5}$, M. Puck Rombach$^{3,4}$,\\ Mason A. Porter$^{3,4}$, Peter J.
Mucha$^{6,7}$, Scott T. Grafton$^{5}$\\
\footnotesize$^1$Department of Physics, University of California, Santa Barbara, CA 93106, USA;\\
\footnotesize$^{2}$ Sage Center for the Study of
the Mind,\\
\footnotesize University of California, Santa Barbara, CA 93106;\\
\footnotesize$^{3}$ Oxford Centre for Industrial and Applied Mathematics,\\
\footnotesize Mathematical Institute, University of Oxford, Oxford OX1 3LB, UK;\\
\footnotesize $^{4}$CABDyN Complexity Centre, University of Oxford, Oxford, OX1 1HP, UK;\\
\footnotesize $^{5}$Department of Psychological and Brain Sciences and UCSB Brain Imaging Center,\\
\footnotesize University of California, Santa Barbara, CA 93106, USA;\\
\footnotesize $^{6}$Carolina Center for Interdisciplinary Applied Mathematics,\\
\footnotesize Department of Mathematics, University of North Carolina,\\
\footnotesize Chapel Hill, NC 27599, USA;\\
\footnotesize $^{7}$Institute for Advanced Materials, Nanoscience \& Technology,\\
\footnotesize University of North Carolina, Chapel Hill, NC 27599, USA;\\
\footnotesize $^*$Corresponding author. Email address: \url{dbassett@physics.ucsb.edu}
}

\maketitle

\newpage


\subsubsection*{Reliability of Temporal Core-Periphery Organization}
\addcontentsline{toc}{subsubsection}{Reliability of Temporal Core-Periphery Organization}

\begin{figure}
\begin{center}
\includegraphics[width=.66\linewidth]{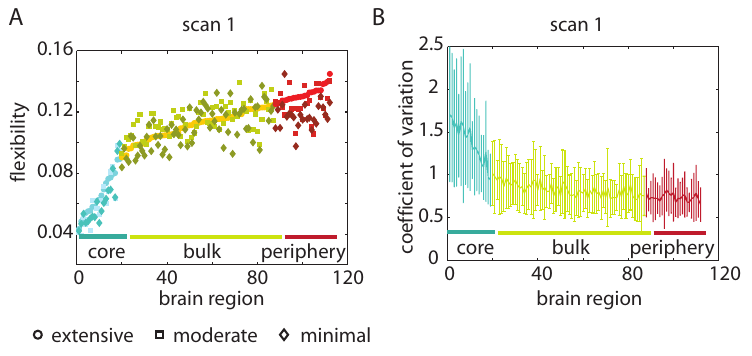}
    \caption[]{\label{fig:cbp1eor} \textbf{Reliability of Temporal Core-Periphery Structure.} Temporal core (cyan), bulk (gold), and periphery (maroon) of dynamic networks determined based on the flexibility of trial blocks in which participants practiced sequences that would eventually be extensively trained. \emph{(A)} Flexibility of the temporal core, bulk, and periphery averaged over the 100 multilayer modularity optimizations and 20 participants for blocks composed of extensively trained (EXT; light circles), moderately trained (MOD; squares), and minimally trained (MIN; dark diamonds) sequences. The darkness of data points indicates scanning session; darker colors indicate earlier scans, so the darkest colors indicate scan 1 and the lightest ones indicate scan 4. \emph{(B)} The coefficient of variation of flexibility calculated over the 100 optimizations and 3 sequence types for all brain regions. Error bars indicate the standard error of the mean CV over participants. Both panels use data from scanning session 1 on day 1 of the experiment (which is prior to home training).
    }
    \end{center}
\end{figure}

A brain region's role in the temporal core, bulk, and periphery is robust across levels of training. Regions identified as part of the core, bulk, or periphery in multilayer networks constructed from the EXT blocks in scanning session 1 have similar flexibilities in the other two levels of training (MOD and MIN; see Fig.~\ref{fig:cbp1eor}A) for the same scanning session. To quantify the variability of a brain region's flexibility, we calculated the coefficient of variation (CV) of flexibility over the 100 optimizations and the 3 levels of training (see Fig.~\ref{fig:cbp1eor}B). The CV is defined as $\mathrm{CV}={\sigma}/{\mu}$, where $\sigma$ is the standard deviation of a given sample and $\mu$ is its mean.  We observe that the variabilities over optimizations and scans (i.e., CV) and over participants (i.e., error bars) are largest in regions designated as part of the temporal core and smallest in regions designated as part of the temporal periphery.

\begin{figure}
\includegraphics[width=1\linewidth]{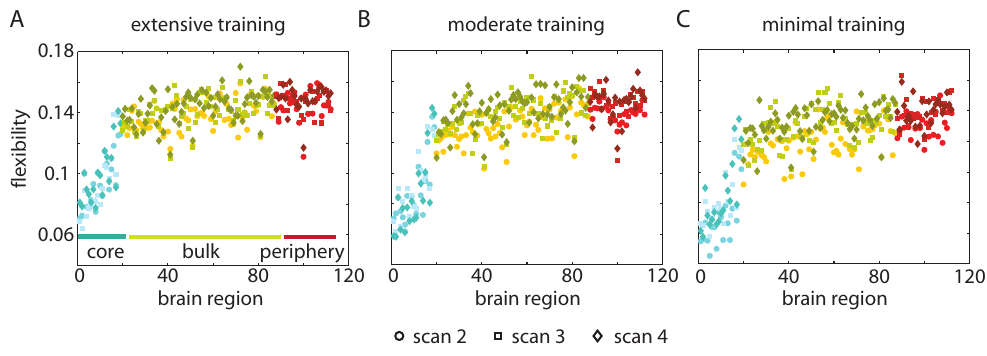}
\begin{center}
    \caption[]{\label{fig:cbp234eor} \textbf{Temporal Core-Periphery Organization Over 42 Days.} Temporal core (cyan), bulk (gold), and periphery (maroon) of dynamic networks defined by trial blocks in which participants practiced sequences that would eventually be \emph{(A)} extensively trained, \emph{(B)} moderately trained, and \emph{(C)} minimally trained for data from scanning sessions 2 (after approximately 2 weeks of training; circles), 3 (after approximately 4 weeks of training; squares), and 4 (after approximately 6 weeks of training; diamonds). The darkness of data points indicates scanning session; darker colors indicate earlier scans, so the darkest colors indicate scan 1 and the lightest ones indicate scan 4.}
    \end{center}
\end{figure}

In addition, regional flexibility is also conserved across both intensity of training (MIN, MOD, and EXT) and duration of training (sessions 1--4). Observe in Fig.~\ref{fig:cbp234eor} that regions identified as part of the temporal core in multilayer networks constructed from the EXT blocks in scanning session 1 exhibit small flexibility for all other scanning sessions and for all 3 training levels (EXT, MOD, and MIN). Regions in the temporal bulk and temporal periphery exhibit a similar amount of flexibility to one another.

\subsubsection*{Reliability of Geometrical Core-Periphery Organization}
\addcontentsline{toc}{subsubsection}{Reliability of Geometrical Core-Periphery Organization}

\begin{figure}
\begin{center}
\includegraphics[width=1\linewidth]{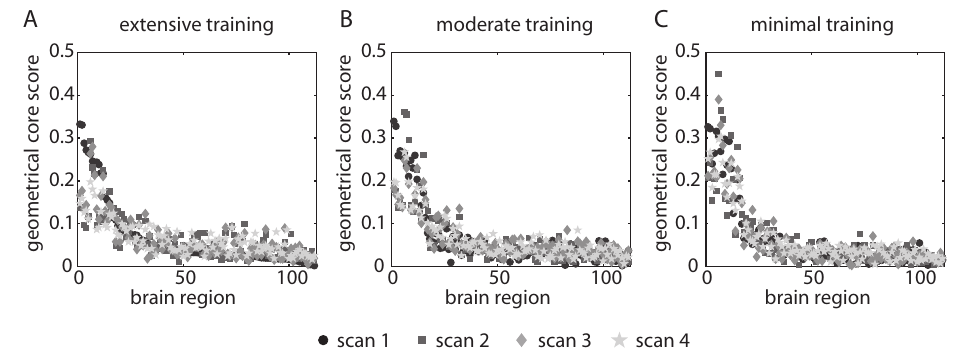}
    \caption[]{\label{fig:sceorreg} \textbf{Geometrical Core-Periphery Organization Over 42 Days.} Geometrical core scores for each brain region defined by the trial blocks in which participants practiced sequences that would eventually be \emph{(A)} extensively trained, \emph{(B)} moderately trained, and \emph{(C)} minimally trained for data from scanning sessions 1 (day 1; black circles), 2 (after approximately 2 weeks of training; dark gray squares), 3 (after approximately 4 weeks of training; gray diamonds), and 4 (after approximately 6 weeks of training; light gray stars). We have averaged the geometrical core scores over blocks and over 20 participants. The order of brain regions is identical for all 3 panels (\emph{A-C}), and we chose this order by ranking regions from high to low geometrical core scores from the EXT blocks on scanning session 1 (on day 1 of the experiment).}
    \end{center}
\end{figure}

As we illustrate in Fig.~\ref{fig:sceorreg}, the geometrical core-periphery organization of the brain was consistent over the 42 days of practice, across sequence types, and throughout variations in the intensity of training (MIN, MOD, and EXT) and in the duration of training (sessions 1--4).

\subsubsection*{Relationship Between Temporal Core-Periphery Organization and Community Structure}
\addcontentsline{toc}{subsubsection}{Relationship Between Temporal Core-Periphery Organization and Community Structure}

\begin{figure}
\begin{center}
\includegraphics[width=0.66\linewidth]{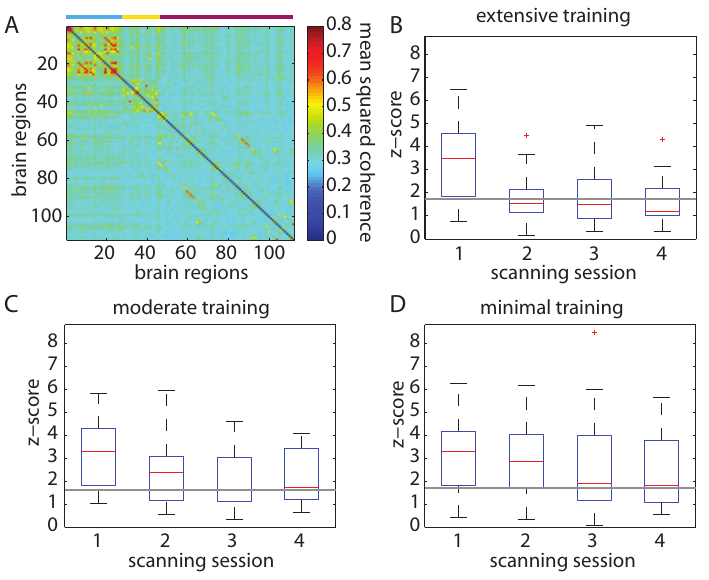}
    \caption[]{\label{fig:modcbp} \textbf{Relationship Between Temporal Core-Periphery Organization and Community Structure.} \emph{(A)} Mean-coherence matrix over all EXT blocks from all participants on scanning day 1. The colored bars above the matrix indicate the 3 communities that we identified from the representative partition. Mean partition similarity z-score $z_{i}$ over all participants for blocks of \emph{(B)} extensively, \emph{(C)} moderately, and \emph{(D)} minimally trained sequences for all 4 scanning sessions over the approximately 6 weeks of training. The horizontal gray lines in panels \emph{(B-D)} indicate the $z_{i}$ value that corresponds to a right-tailed p-value of $0.05$.
    }
    \end{center}
\end{figure}

The division of the brain networks into temporal core, bulk, and peripheral nodes has interesting similarities to their partitioning into communities based on optimizing multilayer modularity. We first noted this similarity when we examined community structure in an object that we call the \emph{mean-coherence matrix}. The mean-coherence matrix $\bar{\mathbf{A}}$ contain elements $\bar{A}_{ij}$ that are equal to the mean coherence between nodes $i$ and $j$ over participants and EXT blocks on day 1 of the experiment. We determined the community structure of this mean-coherence matrix by optimizing the single-layer modularity quality function \cite{NG2004,markfast,Newman2006,Porter2009,Fortunato2010}:
\begin{equation}
	Q_{\mathrm{single-layer}} = \sum_{ij} \left[\bar{A}_{ij} - \frac{k_{i} k_{j}}{2m}\right] \delta(g_{i},g_{j})\,,
\end{equation}
where node $i$ is assigned to community $g_{i}$, node $j$ is assigned to community $g_{j}$, the Kronecker delta $\delta(g_{i},g_{j})=1$ if $g_{i} = g_{j}$ and it equals $0$ otherwise, $k_{i}$ is the strength of node $i$, and $m$ is the mean strength of all nodes in the network. After optimizing this single-layer quality function 100 times, we constructed a representative partition \cite{Bassett2012b} from the set of 100 partitions. (Each partition arises from a single optimization.)  One community in this representative partition, which we show in Fig.~\ref{fig:modcbp}A, appears to have high connectivity to the other two communities: nodes in this first community have edges with strong weights to nodes in the other two communities. This indicates a high coherence in the BOLD time series, and this behavior is consistent with the behavior expected from a network ``core''. A second community in this representative partition appears to have low connectivity to the other two communities: nodes in this community have edges with small weights that connect to nodes in the other two communities. This indicates a low coherence in the BOLD time series, and this behavior is consistent with the behavior expected from a ``periphery''.

It is important to note that we observed this relationship between temporal core-periphery organization and community structure in networks encoded by \emph{mean matrices}. However, networks encoded by \emph{mean matrices} constructed by averaging correlation-based matrices often do not adequately represent the topological or geometrical structure of the ensemble of individual networks from which they are derived \cite{Simpson2012}. We therefore test for a relationship between the temporal core-periphery organization and community structure in the ensemble of networks extracted from individual participants.

A division of the brain into temporal core, bulk, and peripheral regions gives a partition of the functional brain network. We label this partition using the Greek letter $\nu$, and we use the z-score of the Rand coefficient \cite{Traud2010} to test for similarities between this partition and algorithmic partitions, which we label using $\eta$, into communities (based on optimization of multilayer modularity) for each participant, block, and optimization. For each pair of partitions $\nu$ and $\eta$, we calculate the Rand z-score in terms of the total number of node pairs $M$ in the network, the number of pairs $M_{\nu}$ that are in the same community in partition $\nu$ but not in the partition $\eta$, the number of pairs $M_{\eta}$ that are in the same community in partition $\eta$ but not in $\nu$, and the number of node pairs $w_{\nu \eta}$ that are assigned to the same community in both partition $\nu$ and partition $\eta$. The z-score of the Rand coefficient allows one to compare partitions $\eta$ and $\nu$, and it is given by the formula
\begin{equation}
	z_{\nu\eta} = \frac{1}{\sigma_{w_{\nu \eta}}} w_{\nu \eta}-\frac{M_{\nu}M_{\eta}}{M}\,,
\end{equation}
where $\sigma_{w_{\nu \eta}}$ is the standard deviation of $w_{\nu \eta}$. Let the \emph{mean partition similarity} $z^{i}$ denote the mean value of $z_{\nu \eta}$ over all partitions $\eta$ (i.e., for all blocks and all optimizations) for participant $i$.

As we show in Fig.~\ref{fig:modcbp}B-D, we find that communities identified by the optimization of the multilayer modularity quality function (see the ``Materials and Methods'' section in the main manuscript) have significant overlap with the division into temporal core, bulk, and periphery during early learning. The mean values of $z^i$ over participants indicate that there is a significant similarity between the partitions into modules and the partitions into core, bulk, and periphery for networks representing functional connectivity during blocks of extensively, moderately, and minimally trained sequences on scanning day 1. This similarity between community structure and temporal core-periphery organization is also evident for blocks of moderately and minimally trained sequences practiced during later scanning sessions. These results underscore the fact that core-periphery organization can be consistent with community structure.  Note, however, that there is no statistical similarity between partitions into core, bulk, and periphery and partitions into communities for later learning. (As shown in Fig.~\ref{fig:modcbp}B-D, the z-scores for networks that represent the functional connectivity during extensive training in scans 2--4, moderate training in scans 3--4, and minimal training in scan 4 are not significantly greater than expected (i.e., under the null hypothesis of no difference between the partitions).)
Together, this set of results suggests that the relationship between these two types of mesoscale organization can be altered by learning.

\begin{center}
\begin{table*}[ht]
{\normalsize
\hfill{}
\begin{tabular}{|c||c|c|c|c|}
\hline
~ &                     Mean    & Minimum   & Maximum   & Standard Error \\
\hline
\hline
Days  &    ~      & ~         & ~         &   ~           \\
\hline
Between Scans 1 and 2        & 12.00     & 9         & 14        & 0.34          \\
Between Scans 2 and 3        & 12.45     & 10        & 14        & 0.29          \\
Between Scans 3 and 4        & 12.10     & 9         & 22        & 0.63          \\
\hline
\hline
Practice Sessions  &    ~      & ~         & ~         &   ~                    \\
\hline
Between Scans 1 and 2        & 9.70      & 8         & 10        & 0.14          \\
Between Scans 2 and 3        & 9.75      & 4         & 14        & 0.44          \\
Between Scans 3 and 4        & 10.05     & 7         & 13        & 0.32          \\
\hline
\hline
Extensively Trained Trials  &    ~      & ~         & ~         &   ~           \\
\hline
Between Scans 1 and 2        & 620.80     & 512       & 640       & 9.40          \\
Between Scans 2 and 3        & 624.00     & 256       & 896       & 28.57         \\
Between Scans 3 and 4        & 643.20     & 448       & 832       & 20.48         \\
\hline
Moderately Trained Trials  &    ~      & ~         & ~         &   ~           \\
\hline
Between Scans 1 and 2        & 97.00      & 80        & 100       & 1.46          \\
Between Scans 2 and 3        & 97.50      & 40        & 140       & 4.46          \\
Between Scans 3 and 4        & 100.50     & 70        & 130       & 3.20          \\
\hline
Minimally Trained Trials  &    ~      & ~         & ~         &   ~           \\
\hline
Between Scans 1 and 2        & 9.70      & 8         & 10        & 0.14          \\
Between Scans 2 and 3        & 9.75      & 4         & 14        & 0.44          \\
Between Scans 3 and 4        & 10.05     & 7         & 13        & 0.32          \\
\hline
\end{tabular}}
\hfill{}
\caption{\textbf{Experimental Details for Behavioral Data Acquired Between Scanning Sessions.} We give the minimum, mean, maximum, and standard error of the mean over participants for the following variables: the number of days between scanning sessions; the number of practice sessions performed at home between scanning sessions; and the number of trials composed of extensively, moderately, and minimally trained sequences during home practice between scanning sessions. \label{experiment1} }
\end{table*}
\end{center}

\begin{center}
\begin{table*}[ht]
{\normalsize
\hfill{}
\begin{tabular}{|c||c|c|c|c|}
\hline
~ &                     Minimum    & Mean   & Maximum   & Standard Error \\
\hline
\hline
Extensively Trained Blocks                       &    ~      & ~         & ~         &   ~           \\
\hline
During Scan 1               & 6.00      & 9.70      & 10.00     & 0.21          \\
During Scan 2               & 10.00     & 10.00     & 10.00     & 0.00          \\
During Scan 3               & 10.00     & 10.00     & 10.00     & 0.00          \\
During Scan 4               & 8.00      & 9.90      & 10.00     & 0.10          \\
\hline
Moderately Trained Blocks                       &    ~      & ~         & ~         &   ~           \\
\hline
During Scan 1               & 5.00      & 9.70      & 11.00     & 0.27          \\
During Scan 2               & 10.00     & 10.00     & 10.00     & 0.00          \\
During Scan 3               & 10.00     & 10.00     & 10.00     & 0.00          \\
During Scan 4               & 8.00      & 9.90      & 10.00     & 0.10          \\
\hline
Minimally Trained Blocks                       &    ~      & ~         & ~         &   ~           \\
\hline
During Scan 1               & 7.00      & 9.80      & 11.00     & 0.18          \\
During Scan 2               & 10.00     & 10.00     & 10.00     & 0.00          \\
During Scan 3               & 10.00     & 10.00     & 10.00     & 0.00          \\
During Scan 4               & 8.00      & 9.90      & 10.00     & 0.10          \\
\hline
\hline
Length of Extensively Trained Blocks  &    ~      & ~         & ~         &   ~           \\
\hline
During Scan 1               & 52.50     & 61.94     & 72.20     & 1.34          \\
During Scan 2               & 35.50     & 42.36     & 45.90     & 0.72          \\
During Scan 3               & 35.40     & 40.79     & 45.50     & 0.77          \\
During Scan 4               & 34.60     & 40.30     & 45.70     & 0.87          \\
\hline
Length of Moderately Trained Blocks  &    ~      & ~         & ~         &   ~           \\
\hline
During Scan 1               & 50.80     & 61.67     & 72.60     & 1.26          \\
During Scan 2               & 39.70     & 47.56     & 57.20     & 0.80          \\
During Scan 3               & 37.60     & 45.07     & 52.80     & 0.67          \\
During Scan 4               & 37.60     & 43.83     & 50.60     & 0.79          \\
\hline
Length of Minimally Trained Blocks  &    ~      & ~         & ~         &   ~           \\
\hline
During Scan 1               & 52.10     & 61.19     & 70.60     & 1.29          \\
During Scan 2               & 44.10     & 50.02     & 57.70     & 0.73          \\
During Scan 3               & 42.50     & 47.37     & 54.50     & 0.71          \\
During Scan 4               & 39.70     & 45.79     & 54.10     & 0.70          \\
\hline
\end{tabular}}
\hfill{}
\caption{\textbf{Experimental Details for Brain Imaging Data Acquired During Scanning Sessions.} In the top three rows, we give the mean, minimum, maximum, and standard error
over participants for the number of blocks composed of extensively, moderately, and minimally trained sequences during scanning sessions.  In the bottom three rows, we give (in TRs) the mean, minimum, maximum, and standard error of the length over blocks composed of extensively, moderately, and minimally trained sequences during scanning sessions. \label{experiment2}
}
\end{table*}
\end{center}

\small
\begin{table}
\begin{center}
\begin{tabular}{| l | l | l | l |}
\hline
Region Name & Affiliation & Region Name & Affiliation \\
\hline \hline
Frontal pole & \fcolorbox{Goldenrod}{White}{B(R)} \fcolorbox{Goldenrod}{White}{B(L)}  &	 Cingulate gyrus, anterior& \fcolorbox{Goldenrod}{White}{B(R)} \fcolorbox{Goldenrod}{White}{B(L)}   \\
Insular cortex& \fcolorbox{Goldenrod}{White}{B(R)} \fcolorbox{Goldenrod}{White}{B(L)}  &	 Cingulate gyrus, posterior& \fcolorbox{Maroon}{White}{P(R)} \fcolorbox{Goldenrod}{White}{B(L)}   \\
Superior frontal gyrus& \fcolorbox{Goldenrod}{White}{B(R)} \fcolorbox{Goldenrod}{White}{B(L)}    & Precuneus cortex& \fcolorbox{Goldenrod}{White}{B(R)} \fcolorbox{Goldenrod}{White}{B(L)}   \\
Middle frontal gyrus& \fcolorbox{Goldenrod}{White}{B(R)} \fcolorbox{Goldenrod}{White}{B(L)}   & Cuneus cortex& \fcolorbox{SkyBlue}{White}{C(R)} \fcolorbox{SkyBlue}{White}{C(L)}   \\
Inferior frontal gyrus, pars triangularis& \fcolorbox{Goldenrod}{White}{B(R)} \fcolorbox{Maroon}{White}{P(L)}    & Orbital frontal cortex& \fcolorbox{Goldenrod}{White}{B(R)} \fcolorbox{Goldenrod}{White}{B(L)}   \\
Inferior frontal gyrus, pars opercularis& \fcolorbox{Goldenrod}{White}{B(R)} \fcolorbox{Goldenrod}{White}{B(L)}   &Parahippocampal gyrus, anterior& \fcolorbox{Goldenrod}{White}{B(R)} \fcolorbox{Goldenrod}{White}{B(L)}   \\
Precentral gyrus&  \fcolorbox{SkyBlue}{White}{C(R)} \fcolorbox{SkyBlue}{White}{C(L)}  &Parahippocampal gyrus, posterior& \fcolorbox{Goldenrod}{White}{B(R)} \fcolorbox{Maroon}{White}{P(L)}   \\
Temporal pole& \fcolorbox{Goldenrod}{White}{B(R)} \fcolorbox{Goldenrod}{White}{B(L)}   & Lingual gyrus& \fcolorbox{SkyBlue}{White}{C(R)} \fcolorbox{SkyBlue}{White}{C(L)}   \\
Superior temporal gyrus, anterior& \fcolorbox{Goldenrod}{White}{B(R)} \fcolorbox{Goldenrod}{White}{B(L)}   &Temporal fusiform cortex, anterior& \fcolorbox{Maroon}{White}{P(R)} \fcolorbox{Goldenrod}{White}{B(L)}   \\
Superior temporal gyrus, posterior& \fcolorbox{Goldenrod}{White}{B(R)} \fcolorbox{Goldenrod}{White}{B(L)}  & Temporal fusiform cortex, posterior& \fcolorbox{Maroon}{White}{P(R)} \fcolorbox{Maroon}{White}{P(L)}    \\
Middle temporal gyrus, anterior& \fcolorbox{Goldenrod}{White}{B(R)} \fcolorbox{Goldenrod}{White}{B(L)}   &Temporal occipital fusiform cortex&  \fcolorbox{Maroon}{White}{P(R)} \fcolorbox{Maroon}{White}{P(L)}   \\
Middle temporal gyrus, posterior& \fcolorbox{Goldenrod}{White}{B(R)} \fcolorbox{Goldenrod}{White}{B(L)}   &Occipital fusiform gyrus& \fcolorbox{Maroon}{White}{P(R)} \fcolorbox{Maroon}{White}{P(L)}   \\
Middle temporal gyrus, temporooccipital&  \fcolorbox{Maroon}{White}{P(R)} \fcolorbox{Maroon}{White}{P(L)}  & Frontal operculum cortex& \fcolorbox{Goldenrod}{White}{B(R)} \fcolorbox{Maroon}{White}{P(L)}   \\
Inferior temporal gyrus, anterior& \fcolorbox{Goldenrod}{White}{B(R)} \fcolorbox{Goldenrod}{White}{B(L)}   & Central opercular cortex& \fcolorbox{Goldenrod}{White}{B(R)} \fcolorbox{Goldenrod}{White}{B(L)}   \\
Inferior temporal gyrus, posterior& \fcolorbox{Goldenrod}{White}{B(R)} \fcolorbox{Goldenrod}{White}{B(L)} & Parietal operculum cortex& \fcolorbox{Maroon}{White}{P(R)} \fcolorbox{Goldenrod}{White}{B(L)}   \\
Inferior temporal gyrus, temporooccipital& \fcolorbox{Goldenrod}{White}{B(R)} \fcolorbox{Goldenrod}{White}{B(L)} &Planum polare& \fcolorbox{SkyBlue}{White}{C(R)} \fcolorbox{Goldenrod}{White}{B(L)}   \\
Postcentral gyrus& \fcolorbox{Maroon}{White}{P(R)} \fcolorbox{SkyBlue}{White}{C(L)} &Heschl's gyrus& \fcolorbox{SkyBlue}{White}{C(R)} \fcolorbox{SkyBlue}{White}{C(L)}   \\
Superior parietal lobule& \fcolorbox{Goldenrod}{White}{B(R)} \fcolorbox{SkyBlue}{White}{C(L)}	 &Planum temporale& \fcolorbox{Goldenrod}{White}{B(R)} \fcolorbox{Goldenrod}{White}{B(L)}   \\
Supramarginal gyrus, anterior& \fcolorbox{Goldenrod}{White}{B(R)} \fcolorbox{SkyBlue}{White}{C(L)}  &	Supercalcarine cortex& \fcolorbox{SkyBlue}{White}{C(R)} \fcolorbox{SkyBlue}{White}{C(L)}   \\
Supramarginal gyrus, posterior& \fcolorbox{Goldenrod}{White}{B(R)} \fcolorbox{Maroon}{White}{P(L)}  	&Occipital pole& \fcolorbox{SkyBlue}{White}{C(R)} \fcolorbox{Goldenrod}{White}{B(L)}   \\
Angular gyrus& \fcolorbox{Maroon}{White}{P(R)} \fcolorbox{Goldenrod}{White}{B(L)} &Caudate & \fcolorbox{Maroon}{White}{P(R)} \fcolorbox{Goldenrod}{White}{B(L)} \\
Lateral occipital cortex, superior&  \fcolorbox{Maroon}{White}{P(R)} \fcolorbox{Maroon}{White}{P(L)}  	&Putamen& \fcolorbox{Maroon}{White}{P(R)} \fcolorbox{Goldenrod}{White}{B(L)}   \\
Lateral occipital cortex, inferior& \fcolorbox{Maroon}{White}{P(R)} \fcolorbox{Goldenrod}{White}{B(L)} &Globus pallidus& \fcolorbox{Maroon}{White}{P(R)} \fcolorbox{Goldenrod}{White}{B(L)}   \\
Intracalcarine cortex& \fcolorbox{SkyBlue}{White}{C(R)} \fcolorbox{SkyBlue}{White}{C(L)} &Thalamus & \fcolorbox{Maroon}{White}{P(R)} \fcolorbox{Maroon}{White}{P(L)}\\
Frontal medial cortex& \fcolorbox{Goldenrod}{White}{B(R)} \fcolorbox{Goldenrod}{White}{B(L)} &Nucleus Accumbens& \fcolorbox{Goldenrod}{White}{B(R)} \fcolorbox{Goldenrod}{White}{B(L)}   \\
Supplemental motor area& \fcolorbox{SkyBlue}{White}{C(R)} \fcolorbox{SkyBlue}{White}{C(L)}  &	 Parahippocampal gyrus& \fcolorbox{Goldenrod}{White}{B(R)} \fcolorbox{Goldenrod}{White}{B(L)}   \\
Subcallosal cortex& \fcolorbox{Goldenrod}{White}{B(R)} \fcolorbox{Goldenrod}{White}{B(L)}  	 &Hippocampus& \fcolorbox{Goldenrod}{White}{B(R)} \fcolorbox{Goldenrod}{White}{B(L)}   \\
Paracingulate gyrus& \fcolorbox{Goldenrod}{White}{B(R)} \fcolorbox{Goldenrod}{White}{B(L)}  	 &Brainstem& \fcolorbox{Goldenrod}{White}{B(R)} \fcolorbox{Goldenrod}{White}{B(L)}   \\
\hline
\end{tabular}
\end{center}
\caption{\footnotesize \textbf{Brain regions in the Harvard-Oxford (HO) Cortical and Subcortical Parcellation Scheme provided by FSL \cite{Smith2004,Woolrich2009}} and their affiliation to the temporal core (C; cyan), bulk (B; gold), and periphery (P; maroon) for both left (L) and right (R) hemispheres.
}
\label{Table1}
\end{table}
\normalsize


\section*{Methodological Considerations}
\addcontentsline{toc}{section}{Methodological Considerations}


\subsection*{Experimental Factors}
\addcontentsline{toc}{subsection}{Experimental Factors}


\subsubsection*{Effect of Region Size}
\addcontentsline{toc}{subsubsection}{Effect of Region Size}

\begin{figure}
\begin{center}
\includegraphics[width=0.66\linewidth]{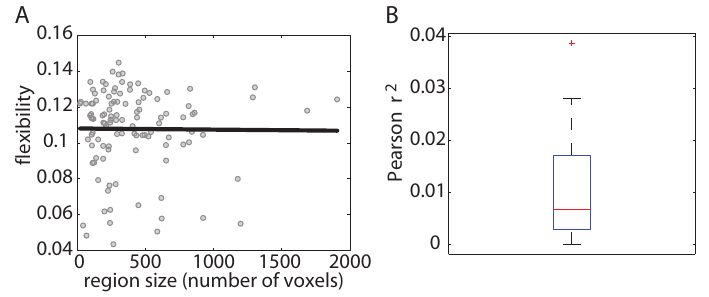}
    \caption[]{\label{fig:regsiz} \textbf{Region Size is Uncorrelated with Flexibility.} \emph{(A)} Scatter plot of the size of the brain region in voxels (averaged over participants) versus the flexibility of the EXT multilayer networks, which we averaged over the 100 multilayer modularity optimizations and the 20 participants. Data points indicate brain regions. The line indicates the best linear fit.  Its Pearson correlation coefficient is $r \doteq -0.009$, and the associated p-value is $p \doteq 0.92$. \emph{(B)} Box plot over the 20 participants of the squared Pearson correlation coefficient $r^2$ between the participant-specific region size in voxels and the participant-specific flexibility averaged over the 100 multilayer modularity optimizations.}
    \end{center}
\end{figure}

Recent studies have noted that brain-region size can affect estimates of hard-wired connectivity strength used in constructing structural connectomes \cite{Hagmann2008,Bassett2010c}. Although the present work is concerned with functional connectomes, it is nevertheless relevant to consider whether or not region size could be a driving effect of the observed core-periphery organization. Importantly, we observe no significant correlation between region size and flexibility (see Fig.~\ref{fig:regsiz}), which suggests that region size is not driving the reported results.

\subsubsection*{Effect of Block Design}
\addcontentsline{toc}{subsubsection}{Effect of Block Design}

Another important factor is the underlying experimental block design and its effect on the correlation structure between brain regions in a single time window (i.e., in a single layer in the multilayer formalism). Two brain regions, such as motor cortex (M1) and supplementary motor area (SMA), might be active during the trial but quiet during the inter-trial interval (ITI). This would lead to a characteristic on-off activity pattern that is highly correlated with all other regions that also turn on with the task and off during the ITI. The frequency of this task-related activity (one on-off cycle per trial, where each trial is of length 4--6 TRs) is included in our frequency band of interest (wavelet scale two, whose frequency range is 0.06--0.12 Hz), and it therefore likely plays a role in the observed correlation patterns between brain regions in a single time window.

Note, however, that our investigations of dynamic network structure---namely, our computations of flexibility of community allegiance---probe functional connectivity dynamics at much larger time scales, and the associated frequencies are an order of magnitude smaller.  They lie in the range 0.0083--0.012 Hz, as there is one time window every 40--60 TRs. At these longer time scales, we can probe the effects of both early learning and extended learning independently of block-design effects.


\subsubsection*{Specificity of Dynamic Network Organization as a Predictor of Learning}
\addcontentsline{toc}{subsubsection}{Specificity of Dynamic Network Organization as a Predictor of Learning}

An important consideration is whether there exist (arguably) simpler properties of brain function than flexibility that could be used to predict learning. We find that the power of activity, the mean connectivity strength, and parameter estimates from a general linear model (GLM) provide less predictive power than flexibility.


\paragraph{Measures of Activity and Connectivity.}
It is far beyond the scope of this study to perform exhaustive computations using all possible measures of brain-region activity, so we focus on two common diagnostics.  One is based on functional connectivity, and the other is based on brain activity. To estimate the strength of functional connectivity, we calculated the mean pairwise coherence between regional wavelet scale-two time series constructed from the BOLD signal, where we took the mean over all possible pairs of regions and all EXT experimental blocks extracted from scans on day 1 for a given subject. To estimate the strength of activity, we calculated the mean signal power of the regional wavelet scale-two time series constructed from the BOLD signal, where we took the mean over all regions and all EXT experimental blocks extracted from scans on day 1 for a given subject. We estimate the power $P_{w_{2}}$ of the wavelet scale-two time series as the square of the time series normalized by its length:
\begin{equation}
	P_{w_{2}} = \sum_{t} \frac{w_{2}(t)^2}{T}\,,
\end{equation}	
where $T$ is the length of the time series \cite{MATLABPowerf,Smith1997}.

We found that neither mean pairwise coherence nor mean power of regional activity measured during the first scanning session could be used to predict learning during the subsequent 10 home training sessions.  For the mean pairwise coherence, we obtained a Pearson correlation of $r \doteq -0.003$ and a p-value of $p \doteq 0.987$.  For the mean power of brain-region activity, we obtained $r \doteq -0.218$ and $p \doteq 0.354$. These results indicate that a prediction similar to that made using the flexibility is not possible using the (arguably) simpler properties of the mean pairwise coherence or the mean power of regional brain activity. They also suggest that the dynamic pattern of coherent functional brain activity is more predictive than means of such activity patterns.


\paragraph{Parameter Estimates for a General Linear Model.}

We determined relative differences in the BOLD signal by using a GLM approach for event-related functional data \cite{GLM,Friston1995}. For each participant, we constructed a single design matrix for event-related fMRI by specifying the onset time and duration of all stimulus events from each scanning session (i.e., the pre-training session and the 3 test sessions). We found estimations of changes in the BOLD signal related to experimental conditions by using the design matrix with the GLM. We modeled the duration of each sequence trial as the time elapsed to produce the entire sequence; in other words, we calculated the movement time (MT), which is a direct measure of the time spent on a task and leads to accurate modeling of BOLD signals using the GLM \cite{Grinband2008}. Separate stimulus vectors indicate each sequence exposure type (EXT, MOD, and MIN) for each scanning session. We took potential differences in brain activity due to rate of movement into account by using the MT for each trial as the modeled duration for the corresponding event.  We convolved events using the canonical hemodynamic response function and temporal derivative. Using the canonical hemodynamic response function (HRF) and its temporal derivative --- we use the implementation in the Statistical Parametric Mapping Toobox (SPM8) [18] --- we then modeled the events that were specified in the stimulus vectors. From this procedure, we obtained a pair of beta images for each event type.  These images correspond to estimates of the HRF and its temporal derivative. Using freely available software \cite{Steffener2010}, we then combined the corresponding beta image pairs for each event type (HRF and its temporal derivative) at the voxel level to form a magnitude image \cite{Calhoun2004}
\begin{equation}
	H = \mathrm{sign}(\hat{B}_{1}) + \sqrt{(\hat{B}_{1}+\hat{B}_{2})}\,,
\end{equation}
where $H$ is called the ``combined amplitude'' of the estimation of the BOLD signal using the HRF ($\hat{B}_{1}$) and its temporal derivative ($\hat{B}_{2}$). \footnote{In this equation, we use the hat notation to indicate that these values are estimated (rather than directly measured) from a general linear model for a response variable (such as regional cerebral blood) at each voxel in a given participant \cite{Friston1995}.} This yielded separate magnitude images for each sequence exposure type (EXT, MOD, and MIN) and session. We then calculated the mean region-based magnitude for each exposure type and session using regions derived from each subject's grey matter-constrained Harvard-Oxford (HO) atlas.

We did not find a significant correlation between the mean parameter estimates averaged over brain regions for the EXT trials in scanning session 1 and learning of the EXT sequences over the subsequent approximately 10 home training sessions.  The Pearson correlation is $r \doteq -0.10$ and the p-value is $p \doteq 0.65$.


\subsubsection*{Subject State-Dependence of Dynamic Network Organization}
\addcontentsline{toc}{subsubsection}{Subject State-Dependence of Dynamic Network Organization}

\rev{Our finding that temporal core-periphery organization predicts the rate of learning across individuals is compelling evidence that the relationship between geometrical and temporal core-periphery organization is related to learning. Nevertheless, it is important to ask whether changes in dynamic community structure and associated mesoscale network organization are related to tasks or to changes in subjects' physiological state over the course of longitudinal imaging \cite{Johnstone2005}. It is clear from studies of behavior, peripheral physiology, and fMRI that subjects can have high levels of anxiety or stress (particularly during their first exposure to MRI) \cite{Lueken2012}. To address this issue, we describe additional evidence that supports our conclusions that the reported changes in dynamic community structure with learning are indeed related to motor tasks.}

\rev{First, we note that we observed temporal and geometrical core-periphery organization consistently over all 4 scanning sessions. In Fig.~\ref{fig:cbp234eor} of the present document, we show that the anatomical identity of nodes in the temporal core, bulk, and periphery are consistent over scanning sessions. In Fig.~\ref{fig:sceorreg} of this document, we show that the anatomical identity of nodes in the geometrical core and periphery are also consistent over scanning sessions. Moreover, Fig.~6 in the main manuscript shows that we observe the relationship between temporal and geometrical core-periphery organization consistently across scanning sessions.}

\rev{Second, we assume that the effects of a subject's mental and physiological state (e.g., anxiety) are greatest during the first imaging session \cite{Chapman2010}. If this is indeed the case, then there could be significant changes of network organization between scans 1 (higher anxiety) and 2 (lower anxiety) that might lead to a spurious interpretation of changes in core-periphery organization. To examine this possibility, we test whether the changes in dynamic community structure and core-periphery organization with learning are robust to the removal of scan 1. Importantly, the trends in Figs.~2 and 5 in the main manuscript remain present if we only examine scans 2--4. We use data from scan 1 for the three box plots located at the point in the horizontal axis at which the number of trials is equal to 50.  (This is the leftmost point of each panel.) See Table 1 in the main manuscript. The 9 box plots located at points on the horizontal axis at which the number of trials is greater than 50 use data from scans 2--4.  Therefore, when we examine only scans 2--4, we still observe a decrease in maximum modularity, an increase in the number of communities, an increase in flexibility, and a decrease in the variance of the geometrical core score with learning.}

\rev{Finally, task-related fMRI BOLD activation magnitude in core, bulk, and peripheral brain regions are not altered significantly across scanning sessions. We employed a repeated-measures analysis of variance (ANOVA) on the training-depth-averaged GLM parameter estimates \cite{Agresti2007}. We treated core, bulk, and periphery designations as categorical factors, and we treated scanning session as a repeated measure. We found a significant main effect (i.e., single-factor effect) of core, bulk, and periphery (an F-statistic \cite{Agresti2007} of $F(2, 38) \doteq 7.88$ and a p-value of $p \doteq 0.00137$) and a non-significant effect of scanning session ($F(3, 57) \doteq 0.615$, $p \doteq 0.584$). These results suggest that a systematic change in the hemodynamic response function across scanning sessions is unlikely to be responsible for the observed learning-related changes in dynamic community structure.}

\rev{Furthermore, we observe that mean GLM parameter estimates in core, bulk, and peripheral brain regions are not correlated significantly with the reported changes in core-periphery structure that accompany learning. The Pearson correlation coefficient between parameter estimates and the variance of the geometrical core score for nodes in the temporal core is $r \doteq 0.20$ (which gives a p-value of $p \doteq 0.52$), for nodes in the temporal bulk is $r \doteq -0.05$ (so $p \doteq 0.86$), and for nodes in the temporal periphery is $r \doteq -0.52$ (so $p \doteq 0.08$). These results provide further evidence that BOLD activation magnitude and dynamic community structure provide distinct insights.}


\subsubsection*{Temporal Core-Periphery Organization and Task-Related Activations}
\addcontentsline{toc}{subsubsection}{Temporal Core-Periphery Organization and Task-Related Activations}

\begin{figure}
\begin{center}
\includegraphics[width=0.99\linewidth]{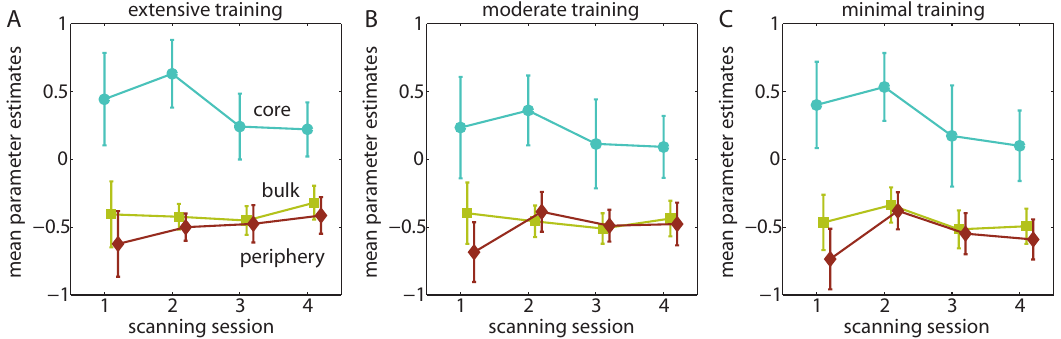}
    \caption[]{\label{fig:glmcbp} \textbf{Temporal Core-Periphery Organization and Task-Related Activations.} Mean GLM parameter estimates for the temporal core (cyan; circles), bulk (gold; squares), and periphery (maroon; diamonds) of dynamic networks defined by the trial blocks in which participants practiced sequences that would eventually be \emph{(A)} extensively trained, \emph{(B)} moderately trained, and \emph{(C)} minimally trained for data from scanning sessions 1 (first day of training),  2 (after approximately 2 weeks of training), 3 (after approximately 4 weeks of training), and 4 (after approximately 6 weeks of training).
    }
    \end{center}
\end{figure}

\rev{One of the strengths of our approach is that we examine the organization of whole-brain functional connectivity and thereby remain sensitive to a wide variety of learning-related changes in the brain that could not be identified using a traditional GLM analysis. Nevertheless, it is useful to explore the relationship between dynamic community structure and task-related activations. In Fig.~\ref{fig:glmcbp}, we show that regions in the temporal core tend to be regions with strong task-related activations, as evinced by high (and positive) values of mean GLM parameter estimates. Conversely, regions in the temporal bulk and periphery tend to lack strong task-related activations, as evinced by low (and negative) values of mean GLM parameter estimates. These results are consistent with our interpretation that the temporal core consists of a small set of regions that are required to perform a given task and that the temporal periphery consists of a set of regions that are associated more peripherally with the task and which are activated in a transient manner.}

\subsection*{Dynamic Community Detection}
\addcontentsline{toc}{subsection}{Dynamic Community Detection}

In the multilayer modularity quality function (see the ``Materials and Methods'' section of the main manuscript), we need to choose values for two parameters \cite{Bassett2012b}: a structural resolution parameter $\gamma$ and a temporal resolution parameter $\omega$. We now examine the effects of these choices on our results.


\subsubsection*{Effect of Structural Resolution Parameter}
\addcontentsline{toc}{subsubsection}{Effect of Structural Resolution Parameter}

In the main manuscript, we used a structural resolution parameter value of $\gamma=1$, which is the most common choice when optimizing the single-layer and multilayer modularity quality functions \cite{Porter2009,Fortunato2010,rb2006}. In this case, ${\bf A}-\gamma {\bf P} = {\bf A}-{\bf P}$, and one is simply subtracting the optimization null model ${\bf P}$ from the adjacency tensor ${\bf A}$. One can decrease $\gamma$ to access community structure at smaller spatial scales (i.e., to examine smaller communities) or increase it to access community structure at larger spatial scales (i.e., to examine larger communities). By examining network diagnostics over a range of $\gamma$ values, we explore the spatial specificity of our results.

\begin{figure}
\begin{center}
\includegraphics[width=0.66\linewidth]{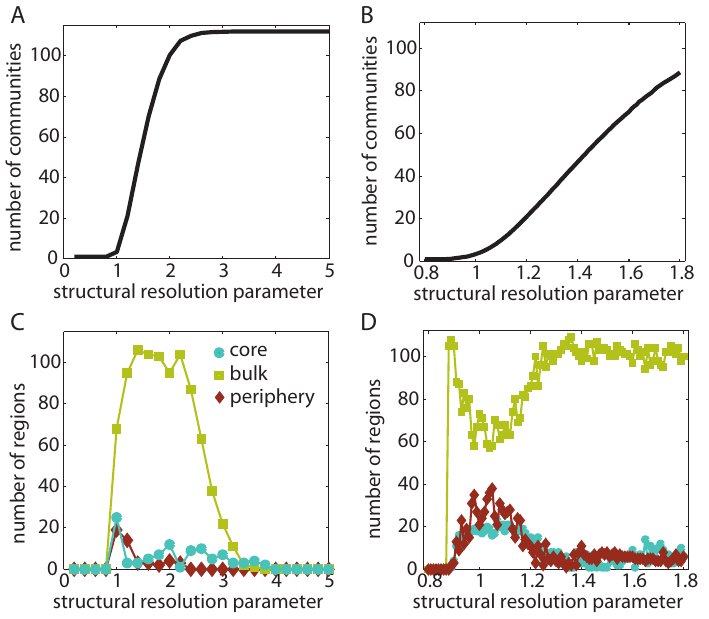}
    \caption[]{\label{fig:gamma} \textbf{(Supplementary Material) Effect of Structural Resolution Parameter.} \emph{(A,B)} Number of communities and \emph{(C,D)} number of regions in the temporal core (cyan; circles), temporal bulk (gold; squares), and temporal periphery (maroon; diamonds) as a function of the structural resolution parameter $\gamma$, where we considered \emph{(A,C)} $\gamma \in [0.2,5]$ in increments of $\Delta \gamma = 0.2$ and \emph{(B,D)} $\gamma \in [0.8,1.8]$ in increments of $\Delta \gamma = 0.01$. We averaged the values in panels \emph{(A)} and \emph{(B)} over 100 multilayer modularity optimizations and over the 20 participants.
    }
    \end{center}
\end{figure}

The mean number of communities in the partitions that we obtained by optimizing multilayer modularity $Q$ varies from the minimum ($1$) to the maximum ($112$) possible value for $\gamma$ approximately in the interval $[0.8,2.5]$ (see Fig.~\ref{fig:gamma}A). We investigate this transition in greater detail in Figs.~\ref{fig:gamma}C,D. Near the value $\gamma=1$, the number of regions in the bulk dips to about $65$, whereas the number of regions in the core and periphery rise to about $20$ and $25$, respectively.  Observe the dip of the bulk curve and bumps of the core and periphery curves in Fig.~\ref{fig:gamma}D.  These features occur for $\gamma$ approximately in the interval $[0.88,1.22]$, which corresponds to partitions that are composed of between approximately $3$ and approximately $20$ communities (with an associated mean community size of between approximately $6$ and approximately $37$ brain regions; see Fig.\ref{fig:gamma}B). This supports our claim that the temporal core-periphery structure that we examine in this study is a genuine mesoscale feature of coherent brain dynamics.


\subsubsection*{Effect of Temporal Resolution Parameter}
\addcontentsline{toc}{subsubsection}{Effect of Temporal Resolution Parameter}

In the main manuscript, we used a temporal resolution parameter value of $\omega=1$. The value $\omega=1$ ensures that the \emph{inter}-layer coupling is equal to the maximum possible value of the \emph{intra}-layer coupling, which we compute from the magnitude-squared coherence (which is constrained to lie in the interval $[0,1]$).  It is important to examine the robustness of results for different values of this parameter, and investigating dynamic network structure at other values of $\omega$ can also provide additional insights \cite{Bassett2012b}. For example, one can decrease $\omega$ to encourage greater variability in community assignments of nodes across individual layers (i.e., across time in temporal networks) or increase it to encourage such community assignments to be more similar across layers. Recall that each node in the temporal multilayer network represents a single brain region at a specified time, and different nodes that represent the same brain region at different times become more likely to be assigned to the same multilayer community as $\omega$ is increased. By examining network diagnostics over a range of $\omega$ values, we can quantify the robustness of our results to differing amounts of temporal variation in community structure.

\begin{figure}
\begin{center}
\includegraphics[width=0.66\linewidth]{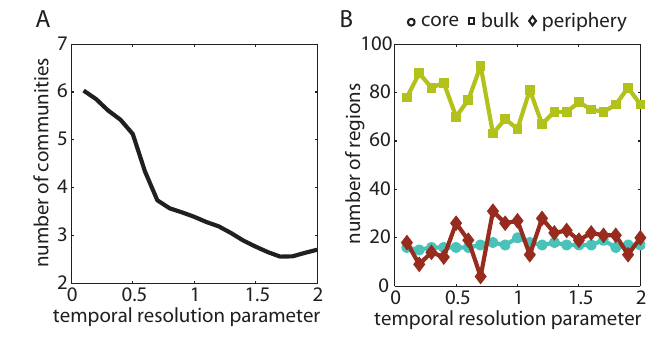}
    \caption[]{\label{fig:omega} \textbf{Effect of Temporal Resolution Parameter.} \emph{(A)} Number of communities averaged over 100 multilayer modularity optimizations and over 20 participants as a function of the temporal resolution parameter $\omega$. \emph{(B)} Number of regions that we identified as part of the temporal core (cyan; circles), temporal bulk (gold; squares), and temporal periphery (maroon; diamonds) as we vary $\omega$ from $0.1$ to $2$ in increments of $\Delta \omega = 0.1$.
    }
    \end{center}
\end{figure}

We varied $\omega$ from 0.1 to 2 in increments of $\Delta \omega = 0.1$. As expected, we find that the number of communities identified in the optimization of the multilayer modularity quality function decreases as $\omega$ is increased (see Fig.~\ref{fig:omega}A). This is consistent with the fact that greater variation of community assignments across time is possible for smaller values of $\omega$. Variation between community assignments of nodes in individual layers can occur in two ways: (1) a small number of regions change community membership from one layer to the next, but the majority of regions retain their community membership; or (2) entire communities lose their identities (via fragmentation, extinction, union, and/or recombination), such that the algorithm identifies either the ``death'' of a community that was present in the previous layer but is not present in the current layer or the ``birth'' of a community that was not present in the previous layer but is present in the current layer.

For each value of $\omega$, we examined the robustness of our division of brain regions into a temporal core, a temporal bulk, and a temporal periphery using the same procedure that we employed for $\omega=1$.  Namely, we defined a temporal core and temporal periphery as those brain regions that were composed, respectively, of the brain regions below and above the 95\% confidence interval of the nodal null model. In Fig.~\ref{fig:omega}B, we report the number of regions in each group as a function of $\omega$.  Interestingly, the number of brain regions that we identified as part of the temporal core varied little over the examined range of $\omega$ values; it remained at approximately $17.0 \pm 1.1$. In fact, $15$ of the $17$ regions that we identified as part of the temporal core at $\omega=1$ were also identified as part of the temporal core at all other values of $\omega$ that we examined.
The number of regions in the temporal bulk and temporal periphery varied more (with values of approximately $75.6 \pm 7.4$ for the bulk and approximately $19.4 \pm 6.8$ for the periphery), which suggests that the separation between the temporal bulk and temporal periphery is less drastic than that between temporal core and temporal bulk. Indeed, the mean flexibility of the core is less similar to the mean flexibility of the bulk than is the latter to the mean flexibility of the periphery. See Fig.~3 of the main manuscript and Figs.~\ref{fig:cbp1eor} and \ref{fig:cbp234eor} of this supplement.

\bibliography{bibfile4}